\documentclass[12pt,preprint]{aastex}
\newcommand{\be}{\begin{equation}}
\newcommand{\ee}{\end{equation}}
\newcommand{\kms}{\,\hbox{km s}^{-1}}
\newcommand{\kpc}{\,\hbox{kpc}}
\newcommand{\pc}{\,\hbox{pc}}
\newcommand{\Mpc}{\,\hbox{Mpc}}

\begin{document}

\title{The slope of the black-hole mass versus velocity dispersion
correlation} 

\author{Scott Tremaine\altaffilmark{1}, Karl Gebhardt\altaffilmark{2}, Ralf
Bender\altaffilmark{3}, Gary Bower\altaffilmark{4}, Alan
Dressler\altaffilmark{5}, S.\ M.\ Faber\altaffilmark{6}, Alexei
V.\ Filippenko\altaffilmark{7}, Richard Green\altaffilmark{8}, Carl
Grillmair\altaffilmark{9}, Luis C.\ Ho\altaffilmark{5}, John
Kormendy\altaffilmark{2}, Tod R.\ Lauer\altaffilmark{8}, John
Magorrian\altaffilmark{10}, Jason Pinkney\altaffilmark{11}, and Douglas
Richstone\altaffilmark{11}}

\altaffiltext{1}{Princeton University Observatory, Peyton Hall,
Princeton, NJ 08544; tremaine@astro.princeton.edu}

\altaffiltext{2}{Department of Astronomy, University of Texas, 
RLM 15.308, Austin, Texas 78712; gebhardt@astro.as.utexas.edu,
kormendy@astro.as.utexas.edu}

\altaffiltext{3}{Universit\"ats-Sternwarte, Scheinerstra\ss e 1,
M\"unchen 81679, Germany; bender@usm.uni-muenchen.de}

\altaffiltext{4}{Computer Sciences Corporation, Space Telescope Science
Institute, 3700 San Martin Drive, Baltimore, MD 21218; bower@stsci.edu}

\altaffiltext{5}{The Observatories of the Carnegie Institution of
Washington, 813 Santa Barbara St., Pasadena, CA 91101;
dressler@ociw.edu, lho@ociw.edu}

\altaffiltext{6}{UCO/Lick Observatories, University of California,
Santa Cruz, CA 95064; faber@ucolick.org}

\altaffiltext{7}{Department of Astronomy, University of California,
Berkeley, CA 94720-3411; alex@astro.berkeley.edu}

\altaffiltext{8}{National Optical Astronomy Observatories, P. O. Box
26732, Tucson, AZ 85726; green@noao.edu, lauer@noao.edu}

\altaffiltext{9}{SIRTF Science Center, Mail Stop 220-6, 1200 East California
Blvd., Pasadena, CA 91125; carl@ipac.caltech.edu}

\altaffiltext{10}{Department of Physics, University of Durham, Rochester
Building, Science Laboratories, South Road, Durham DH1 3LE, UK;
John.Magorrian@durham.ac.uk}

\altaffiltext{11}{Dept. of Astronomy, Dennison Bldg., Univ. of
Michigan, Ann Arbor 48109; jpinkney@\-astro.\-lsa.\-umich.\-edu,
dor@astro.lsa.umich.edu}
 
\begin{abstract}
Observations of nearby galaxies reveal a strong correlation between the mass
of the central dark object $M_\bullet$ and the velocity dispersion $\sigma$ of
the host galaxy, of the form $\log(M_\bullet/M_\odot)=\alpha+
\beta\log(\sigma/\sigma_0)$; however, published estimates of the slope $\beta$
span a wide range (3.75 to 5.3). Merritt \& Ferrarese have argued that low
slopes ($\lesssim 4$) arise because of neglect of random measurement errors in
the dispersions and an incorrect choice for the dispersion of the Milky Way
Galaxy. We show that these explanations, and several others, account for at
most a small part of the slope range. Instead, the range of slopes arises
mostly because of systematic differences in the velocity dispersions used by
different groups for the same galaxies. The origin of these differences
remains unclear, but we suggest that one significant component of the
difference results from Ferrarese \& Merritt's extrapolation of central
velocity dispersions to $r_e/8$ ($r_e$ is the effective radius) using an
empirical formula. Another component may arise from dispersion-dependent
systematic errors in the measurements. A new determination of the slope using
31 galaxies yields $\beta=4.02\pm0.32$, $\alpha=8.13\pm0.06$, for
$\sigma_0=200\kms$. The $M_\bullet$--$\sigma$ relation has an intrinsic
dispersion in $\log M_\bullet$ that is no larger than 0.3 dex, and may be
smaller if observational errors have been underestimated. In an Appendix, we
present a simple kinematic model for the velocity-dispersion profile of the
Galactic bulge.

\end{abstract}

\section{Introduction}

Observations of the centers of nearby early-type galaxies (ellipticals,
lenticulars, and spiral bulges) show that most or all contain massive dark
objects (hereafter ``black holes''). The masses of these objects are
consistent with the density of quasar remnants expected from energy arguments
\citep{sol82,fab99,yu02}. There appears to be a strong correlation between
the mass $M_\bullet$ of the black hole and the velocity dispersion $\sigma$ of
the host galaxy, of the form\footnote{All logarithms in this paper are base
10.} 
\be
\log (M_\bullet/M_\odot)=\alpha+\beta\log(\sigma/\sigma_0),
\label{eq:gebrel}
\ee
where $\sigma_0$ is some reference value (here chosen to be
$\sigma_0=200\kms$). The first published estimates of the slope $\beta$,
$5.27\pm0.40$ \citep{FM00a} and $3.75\pm0.3$ \citep{G00}, differed by 3
standard deviations. Subsequently, Ferrarese \& Merritt (hereafter FM) revised
their slope downwards, to $4.8\pm0.5$ \citep{FM00b}, $4.72\pm0.36$
\citep{MF01a}, $4.65\pm 0.48$ \citep{MF01b}, and then $4.58\pm 0.52$
\citep{fer02}. Although the discrepancy between the estimate by Gebhardt et
al. (hereafter the Nukers) and the estimates by FM has declined monotonically
with time, and is now only 1.4 standard deviations, it is still worthwhile to
understand the reasons behind it. In particular, the slope is the most
important point of comparison to theoretical models that attempt to explain
the $M_\bullet$--$\sigma$ relation \citep{ada01,bur01,hae00,ost00}.

This paper has three main goals. (i) In \S\S\ref{sec:fit}--\ref{sec:why} we
explore the reasons for the wide range in estimated slopes of the
$M_\bullet$--$\sigma$ relation. In \S\ref{sec:fit} we focus on the statistical
techniques used to estimate slopes by the two groups; we shall argue that the
estimator used by the Nukers is more accurate, but that the choice of
estimator cannot explain most of the differences in slope between FM and the
Nukers.  In \S\ref{sec:data} we describe the data sets used by the two
groups. In \S\ref{sec:why} we examine several explanations that have been
proposed for the slope range, including the neglect of random measurement
errors in the dispersions, the dispersion used for the Milky Way, and
differences in sample selection, and show that none of these is viable. We
argue instead that the slope range reflects systematic differences in
the velocity dispersion measurements used by the two groups.  (ii) In
\S\ref{sec:new} we present a new analysis of the $M_\bullet$--$\sigma$
relation using recent data. (iii) Finally, in the Appendix we model the
velocity-dispersion profile of the Milky Way bulge, which helps to fix the
low-mass end of the $M_\bullet$--$\sigma$ relation.

\section{The fitting algorithm} 

\label{sec:fit}

The data consist of $N$ galaxies with measured black-hole masses, velocity
dispersions, and associated uncertainties. We assume that there is an
underlying relation of the form
\be
y=\alpha +\beta x,
\label{eq:regress}
\ee
where $y=\log(M_\bullet/M_\odot)$, $x=\log(\sigma/\sigma_0)$. We
assume that the measurement errors are symmetric in $x$ and $y$ with
root-mean-square (rms) values $\epsilon_{xi}$ and $\epsilon_{yi}$ for galaxy
$i$. The goal is to estimate the best-fit values of $\alpha$ and $\beta$ and
their associated uncertainties.

The Nukers and FM use two quite different estimators. In this section we
review the assumptions inherent in the two estimators and their respective
advantages and disadvantages. In subsequent sections we shall usually give
results for both estimators; we shall find that the differences are
significant, but not large enough to explain the slope range.

The Nukers' estimate is based on minimizing 
\be 
\chi^2\equiv \sum_{i=1}^N{(y_i-\alpha - \beta x_i)^2\over
\epsilon_{yi}^2+\beta^2\epsilon_{xi}^2}, 
\label{eq:chisq}
\ee
(e.g., Press et al.\ 1992, whose procedures we use). The ``1$\sigma$''
uncertainties in $\alpha$ and $\beta$ are given by the maximum range of
$\alpha$ and $\beta$ for which $\chi^2-\chi^2_{\rm min}\le 1$. An attractive
feature of this approach is that the variables $x$ and $y$ are treated
symmetrically; in other words, if we set $\widetilde\beta=1/\beta$,
$\widetilde\alpha=-\alpha/\beta$, equation (\ref{eq:chisq}) can be rewritten
in the form
\be 
\chi^2\equiv \sum_{i=1}^N{(x_i - \widetilde\alpha - \widetilde\beta y_i)^2\over
\epsilon_{xi}^2+{\widetilde\beta}^2\epsilon_{yi}^2},
\label{eq:chisqinv}
\ee 
which has the same form as equation (\ref{eq:chisq}) if $x\leftrightarrow y$,
$\alpha\leftrightarrow \widetilde\alpha$, and $\beta\leftrightarrow
\widetilde\beta$. 
This symmetry ensures that we are not assuming (for example) that $y$ is the
dependent variable and $x$ is the independent variable in the correlation;
this agnosticism is important because we do not understand the physical
mechanism that links black-hole mass to dispersion. We shall call estimators
of this kind ``$\chi^2$ estimators'' and denote them by $\alpha_\chi$,
$\beta_\chi$.

One limitation is that this approach does not account for any intrinsic
dispersion in the $M_\bullet$--$\sigma$ relation (i.e., dispersion due to the
galaxies themselves rather than to measurement errors). Thus, for example, one
or two very precise measurements with small values of $\epsilon_{xi}$ and
$\epsilon_{yi}$ can dominate $\chi^2$, even though the large weight given to
these observations is unrealistic if the intrinsic dispersion is larger than
the measurement errors. There are two heuristic approaches that address this
concern. (i) Simply set $\epsilon_{yi}\equiv\epsilon_y=\hbox{constant}$,
corresponding to the same fractional uncertainty in all the black-hole mass
estimates. The value of $\epsilon_y$ is adjusted so that the value of $\chi^2$
per degree of freedom is equal to its expectation value of unity. This
approach was adopted by \citet{G00}.  (ii) Replace $\epsilon_{yi}$ by
$(\epsilon_{yi}^2+\epsilon_0^2)^{1/2}$, where the unknown constant
$\epsilon_0$, which represents the intrinsic dispersion, is adjusted so that
the value of $\chi^2$ per degree of freedom is unity. The second procedure is
preferable if and only if the individual error estimates $\epsilon_{yi}$ are
reliable. We shall use both approaches in \S\ref{sec:slope}.

FM use the estimator
\be
\beta_{\rm AB}={\sum_{i=1}^N(y_i-\langle y\rangle)(x_i-\langle x\rangle)\over\sum_{i=1}^N
(x_i-\langle x\rangle)^2-\sum_{i=1}^N\epsilon_{xi}^2}, \qquad \alpha_{\rm
AB}=\langle y\rangle -\beta_{\rm AB} \langle x\rangle;
\label{eq:ab}
\ee
here $\langle x\rangle\equiv N^{-1}\sum_{i=1}^N x_i$ and $\langle
y\rangle\equiv N^{-1}\sum_{i=1}^N y_i$ are the sample means of the two
variables. This estimator is described by \citet{ab96}, who also provide
formulae for the uncertainties in $\alpha$ and $\beta$. The Akritas-Bershady
(hereafter AB) estimator accounts for measurement uncertainties in both
variables, and is asymptotically normal and consistent. When $\epsilon_{xi}=0$
and $\epsilon_{yi}\equiv\epsilon_y=\hbox{constant}$, the AB and $\chi^2$
estimators give the same estimates for $\alpha$ and $\beta$ (but not their
uncertainties).

\begin{figure}
\plotone{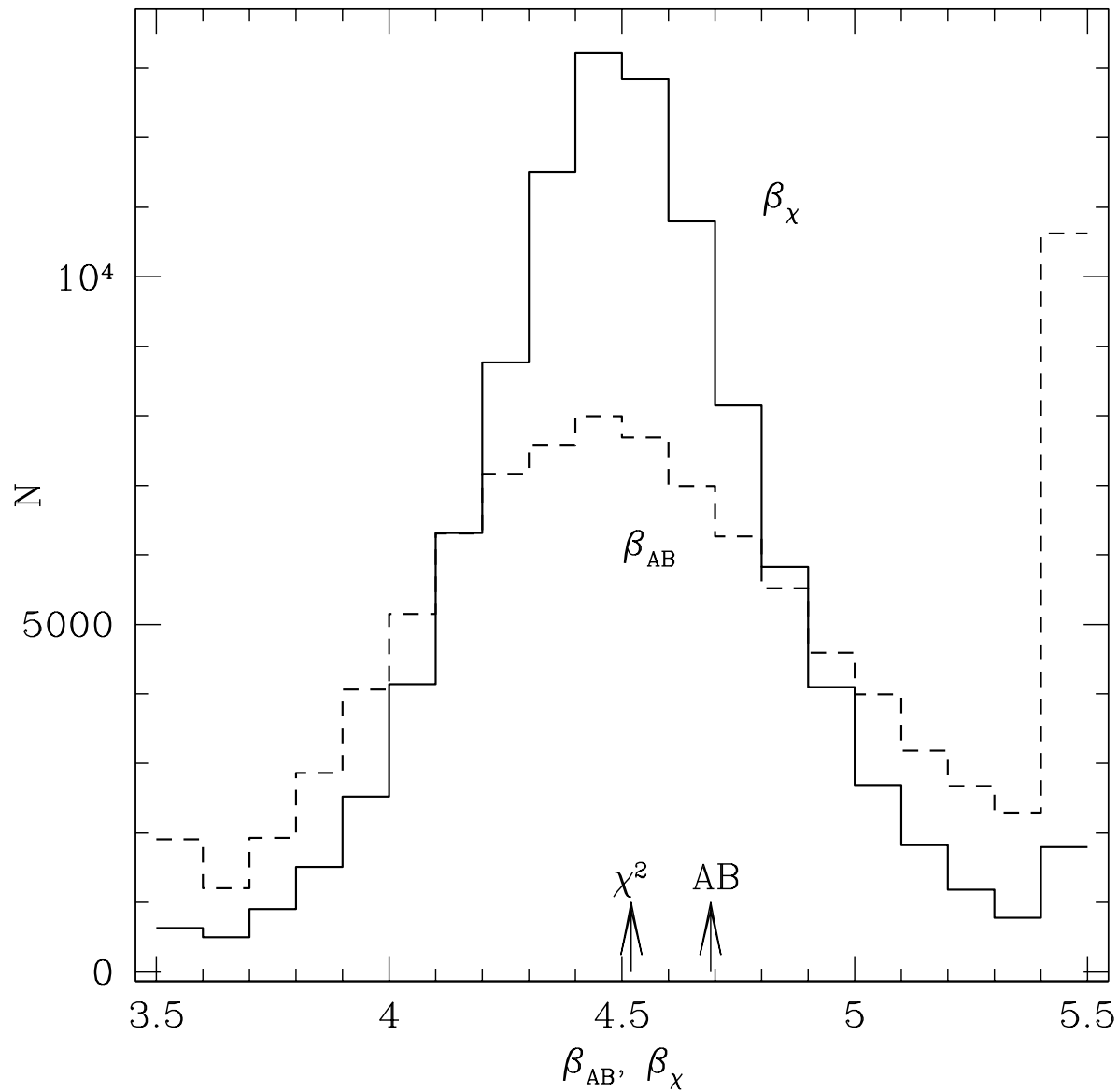}
\caption{The distribution of the estimators $\beta_\chi$ (eq.\ \ref{eq:chisq}; 
solid line) and $\beta_{\rm AB}$ (eq.\ \ref{eq:ab}; dashed line) for 100,000
Monte Carlo simulations of a sample with $\beta=5$ that resembles the actual
sample FM1 (12 galaxies, distributed as a Gaussian with standard deviation
0.20 in $x$, and Gaussian measurement errors with standard deviations
$\epsilon_x=0.06$, $\epsilon_y=0.18$). Values $>5.5$ or $<3.3$ are plotted in
the outermost bins of the histogram. The sample means are marked by arrows.}
\label{fig:hist}
\end{figure}

Despite its merits, the AB estimator has several unsettling properties. (i)
The measurement errors in velocity dispersion, $\epsilon_{xi}$, only enter
equation (\ref{eq:ab}) through the sum $\sum\epsilon_{xi}^2$. Thus, for
example, a single low-precision measurement can dominate both
$\sum\epsilon_{xi}^2$ and $\sum_{i=1}^N (x_i-\langle x\rangle)^2$, rendering
the estimator useless, no matter how many high-precision measurements are in
the sample. (ii) The errors in the black-hole mass determinations
$\epsilon_{yi}$ do not enter equation (\ref{eq:ab}) at all: all observations
are given equal weight, even if some are known to be much less precise than
others. (iii) We have argued above that the variables $x$ and $y$ should be
treated symmetrically, but this is not the case in equation (\ref{eq:ab}).
(iv) Even if the variables $x_i$ are drawn from a Gaussian distribution, there
will occasionally be samples for which the denominator of equation
(\ref{eq:ab}) is near zero. In this case the estimator $\beta_{\rm AB}$ will
be very large. These occasional large excursions are frequent enough that the
variance of $\beta_{\rm AB}$ in a population of galaxy samples is infinite, no
matter how large the number $N$ of data points may be. (v) Figure
\ref{fig:hist} shows the distribution of estimates of $\beta_\chi$ (solid
line) and $\beta_{\rm AB}$ (dashed line) obtained from 100,000 Monte Carlo
trials drawn from a population that has $\beta=4.5$ and other parameters
similar to the sample FM1 defined below (for details see figure caption).  The
distribution of $\beta_\chi$ is substantially narrower than $\beta_{\rm AB}$
(note that values of either estimator outside the range of the histogram are
plotted in the outermost bins). The estimator $\beta_\chi$ has a mean of
$4.52$ and a standard deviation of $0.36$. The distribution of $\beta_{\rm
AB}$ has a mean of $4.69$, and as stated above the standard deviation of this
mean is infinite.  Thus, in this example at least, $\beta_{\rm AB}$ is both
biased and inefficient. 

In this paper, we sometimes use a third fitting procedure, which is closely
related to principal component analysis \citep{ken83}. Suppose that the
intrinsic distribution of $x$ and $y$ (the distribution that would be observed
in the absence of measurement errors) is a biaxial Gaussian, with major and
minor axes having standard deviations $\sigma_a$ and $\sigma_b$, respectively,
and the major axis having slope $\beta\equiv\tan\theta$. If $\sigma_b$ were
zero, all of the points would lie exactly on a line of slope $\beta$; thus
$\sigma_b$ characterizes the intrinsic dispersion in the correlation between
$x$ and $y$. Let us also assume that the measurement errors are Gaussian, with
standard deviations $\epsilon_x$ and $\epsilon_y$ that are the same for all
galaxies.  The observed distribution of $x$ and $y$, which is obtained by
convolving the intrinsic Gaussian with the measurement errors, is still
Gaussian. The shape of this Gaussian is fully described by the three
independent components of the symmetric $2\times2$ dispersion tensor
\be
\sigma_{ij}\equiv\langle(x_i-\langle x_i\rangle)(x_j-\langle
x_j\rangle)\rangle, \qquad i=1,2,\ j=1,2,
\ee
where $(x_1,x_2)\equiv (x,y)$ and $\langle\cdot\rangle$ denotes a sample
average. In this idealized but plausible model, at most three of the five
parameters $\epsilon_x$, $\epsilon_y$, $\sigma_a$, $\sigma_b$, and $\beta$ can
be determined from the data, no matter how many galaxies we observe. For
example, if $\epsilon_x$ and $\epsilon_y$ are known, the other parameters can
be estimated using the formulae
\begin{eqnarray}
\tan2\theta &=& {2\sigma_{xy}\over
\sigma_{xx}-\sigma_{yy}+\epsilon_y^2-\epsilon_x^2}, \nonumber \\
\sigma_b^2 &=& \sigma_{xx}-\epsilon_x^2-\sigma_{xy}\cot\theta=
\sigma_{yy}-\epsilon_y^2-\sigma_{xy}\tan\theta, \nonumber \\
\sigma_a^2 &=& \sigma_{xx}-\epsilon_x^2+\sigma_{xy}\tan\theta=
\sigma_{yy}-\epsilon_y^2+\sigma_{xy}\cot\theta;
\label{eq:gauss}
\end{eqnarray}
there are two solutions for $\theta$ differing by $\pi/2$, and we choose the
solution for which $\sigma_a>\sigma_b>0$. These equations, which we call
Gaussian estimators, are related to the AB estimator (\ref{eq:ab}), which in
this notation is simply
$\tan\theta=\sigma_{xy}/(\sigma_{xx}-\epsilon_x^2)$. However, the Gaussian
estimators have the advantages that (i) they are symmetric in $x$ and $y$, and
(ii) they account naturally for the possibility that there is an intrinsic
dispersion $\sigma_b$ in the $M_\bullet$--$\sigma$ correlation. The Gaussian
estimators can easily be extended to include measurement errors that differ
from galaxy to galaxy, and to provide uncertainties in the estimators (e.g.,
Gull 1989), and with these extensions they are likely to provide a more
reliable slope estimator than either the $\chi^2$ or AB estimators.

We close this section with a general comment on fitting linear relations such
as (\ref{eq:gebrel}). The choice of the reference value $\sigma_0$ affects the
uncertainty in $\alpha$ and the covariance between the estimated values of
$\alpha$ and $\beta$. A rough rule of thumb is that $\sigma_0$ should be
chosen near the middle of the range of values of $\sigma$ in the galaxy sample
to minimize the uncertainty in $\alpha$ and the correlation between $\alpha$
and $\beta$. As an example, \citet{FM00b} use $\sigma_0=1\kms$ and find an
uncertainty in $\alpha$ of $\pm1.3$. However, most of this uncertainty arises
because errors in $\alpha$ and $\beta$ are strongly correlated at this value
of $\sigma_0$ (correlation coefficient $r=-0.998$). Simply by choosing
$\sigma_0=200\kms$ the uncertainty in $\alpha$ is reduced by a factor of more
than ten, to $\pm0.09$.

\section{The data}

\label{sec:data}

The $M_\bullet$--$\sigma$ relation has been explored in the literature using a
number of distinct datasets:

\begin{enumerate}

\item {\bf Sample FM1} \quad Much of FM's analysis is based on a set of 12
galaxies with ``secure'' black-hole mass estimates (sample A, Table 1 of
Ferrarese \& Merritt 2000b). However, their definition of ``secure'' is not
itself secure: in \S\ref{sec:new}, we reject one of the galaxies in this
sample (NGC 4374) because of concerns about the reliability of its mass
estimate, and the best estimate of the mass of another (IC 1459) has recently
increased by a factor of six. Half of the black-hole mass estimates in this
sample come from gas kinematics, as determined by {\it Hubble Space Telescope}
({\it HST}\,) emission-line spectra, and the remainder from stellar and maser
kinematics. Unless otherwise indicated, when discussing this sample we shall
use the upper and lower limits to the dispersion and black-hole mass given by
\citet{FM00b}.\footnote{The error bars in $x$ and $y$ are
given by $(\log\sigma_{\rm upper}-\log\sigma_{\rm lower})/2$ and
$(\log M_{\bullet,\rm upper}-\log M_{\bullet,\rm lower})/2$,
respectively.}\ The slope estimators then yield
\be
\beta_\chi=4.47\pm0.44, \qquad    \beta_{\rm AB}=4.81\pm0.55.
\ee
The minimum $\chi^2$ per degree of freedom is 0.69, which indicates an
acceptable fit; thus there is no evidence for any intrinsic dispersion in this
sample. 

\item {\bf Sample G1} \quad The sample used by \citet{G00} contains 26 
galaxies. Of these, the majority (18) of the mass estimates are from
axisymmetric dynamical models of the stellar distribution function, based on
{\it HST} and ground-based absorption-line spectra.  All of the galaxies in
sample FM1 are contained in this sample except for NGC 3115. The stated rms
fractional uncertainty in the black-hole masses is 0.22 dex, but following
\citet{G00}, we shall adopt $\epsilon_y=0.30$, which yields a minimum $\chi^2$
per degree of freedom equal to unity. \citet{G00} take $\epsilon_x=0$,
corresponding to negligible uncertainties in the dispersions; this
approximation is discussed in \S\ref{sec:disperr}. The slope estimators then
yield
\be
\beta_\chi=3.74\pm 0.30, \qquad \beta_{\rm AB}=3.74\pm 0.23.
\ee
A maximum-likelihood estimate of the intrinsic dispersion in black-hole mass
at constant velocity dispersion for this sample is $0.22\pm0.05$ dex. 

\item {\bf Sample FM2} \quad \citet{MF01b} supplement sample FM1 with
10 additional galaxies, mostly taken from \citet{kg01}, for a total of 22
galaxies.  The stated rms fractional uncertainty in the black-hole masses is
0.24 dex. The slope estimators yield
\be
\beta_\chi=4.78\pm 0.43, \qquad \beta_{\rm AB}=4.65\pm 0.49.
\label{eq:mfb}
\ee
The minimum $\chi^2$ per degree of freedom is 1.1, and there is no evidence
for any intrinsic dispersion in the black-hole mass. 

\item {\bf Sample G2} \quad These are the 22 galaxies listed by \citet{kg01}
that are also in sample FM2. By comparing samples FM2 and G2 we can isolate
the effects of different treatments of the same galaxies. We shall assume 20\%
uncertainty in the dispersion of the Milky Way, and 5\% uncertainty in the
velocity dispersions of external galaxies (see \S\S\ref{sec:disperr} and
\ref{sec:mw}). Using G2's stated uncertainties in the black-hole masses, the
slope estimators yield $\beta_\chi=3.70 \pm 0.20$, $\beta_{\rm AB}=3.61\pm
0.31$. The minimum $\chi^2$ per degree of freedom is 2.8, which suggests that
either the uncertainties in the black-hole masses are underestimated or there
is an intrinsic dispersion in black-hole mass. Adding an intrinsic dispersion
of 0.17 dex decreases the value of $\chi^2$ per degree of freedom to unity,
and reduces the best-fit slope to
\be
\beta_\chi=3.61 \pm 0.29, \qquad \beta_{\rm AB}=3.61\pm 0.31.
\label{eq:kg}
\ee
A maximum-likelihood estimate of the intrinsic dispersion in black-hole mass
at constant velocity dispersion for this sample is $0.16\pm0.05$ dex. 

\end{enumerate}

\section{Why are the slopes different?}

\label{sec:why}

Our goal is to determine why different investigations yield such a wide range
of slopes. In particular, the two samples from FM give slopes $\gtrsim 4.5$
(``high'' slopes) with both the $\chi^2$ and AB estimators, while the two
samples from the Nukers give slopes $\lesssim 4.0$ (``low'' slopes) with both
estimators. In \S\S\ref{sec:disperr}--\ref{sec:samp} we describe several
explanations for the slope range that have been proposed in the literature,
all of which are found to be inadequate. In \S\ref{sec:apdisp} we suggest that
systematic differences in the dispersions used by FM and the Nukers are
responsible for most of the slope discrepancy. 

\begin{figure}
\plotone{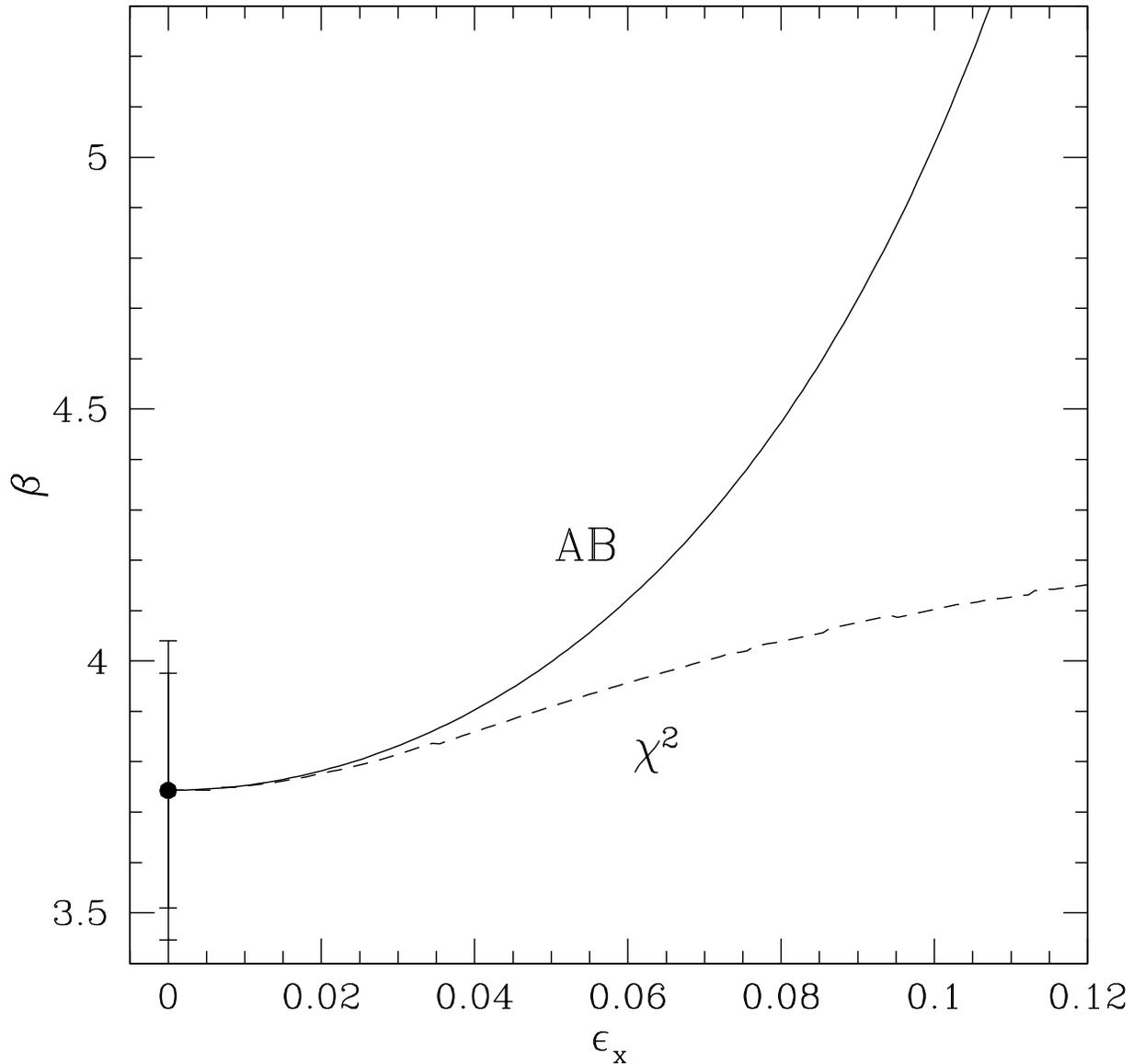}
\caption{The dependence of the slope $\beta$ on the assumed measurement
uncertainty in velocity dispersion, for the sample G1. The abscissa is the rms
measurement error in $\log\sigma$. Solid and dashed lines show the slopes
derived from the AB and $\chi^2$ estimators, respectively. The error bars show
the computed uncertainty in the slope at zero error. The formal uncertainty in
the dispersion measurements of G1 is $\epsilon_x\simeq 0.01$; allowing for
possible systematic errors in the stellar template and continuum subtraction
increases $\epsilon_x$ to $\sim 0.02$ (5\%).}
\label{fig:err}
\end{figure}

\subsection{Measurement errors in velocity dispersion}

\label{sec:disperr} 

\citet{MF01a} argue that random measurement errors in the velocity dispersion
can have a significant effect on the slope of the $M_\bullet$--$\sigma$
regression.  In particular, they claim that the Nukers' assumption of zero
measurement error in $\sigma$ leads them to underestimate the slope. To test
this claim, we plot in Figure \ref{fig:err} the slope $\beta$ derived from the
G1 sample using both the AB and $\chi^2$ estimators, as a function of the
assumed rms error $\epsilon_x$ in the log of the velocity dispersion.

For nearly all of the galaxies in sample G1, the data typically have
signal-to-noise ratio around 100, and the formal uncertainties in the
dispersions are around 2--3\% ($\epsilon_x=0.009$--0.013). However, at this
level, stellar template variations, assumptions about the continuum shape,
fitting regions used, and atmospheric seeing conditions all can have a
noticeable effect on the estimated dispersion. To account crudely for these
systematic errors, we double the uncertainties in the dispersions, to 5\%
($\epsilon_x=0.021$). The uncertainty is larger in the Milky Way (see
\S\ref{sec:mw}), and in a few galaxies that we have not observed ourselves and
that do not have accurate dispersion profiles in the literature.  The
statement of \citet{MF01a} that velocity-dispersion errors are ``easily at the
10\% level'' is indeed correct for the sample FM1, where the rms fractional
error in the dispersions is 14\% ($\epsilon_x=0.057$), but their dispersions
are mostly based on heterogeneous data that are 20--30 years old
\citep{dav87}.

Figure \ref{fig:err} shows that the effect of random errors in the dispersions
is negligible: at the 5\% level, the change in $\beta$ for sample G1 is
only 0.03 or 0.04 for the $\chi^2$ and AB estimators respectively, and even at
the 10\% level the corresponding changes are only 0.12 and 0.16. 

\subsection{Measurement errors in black-hole mass}

We next ask whether the combined effects of varying assumptions about
measurement errors in both velocity dispersion and black-hole mass can explain
the discrepancy between the high and low slopes. As usual, we parametrize
these uncertainties by $\epsilon_x$ and $\epsilon_y$, the rms measurement
error in the log of the velocity dispersion and black-hole mass. For
simplicity, in this subsection these errors are assumed to be the same for all
galaxies in each sample. The effects of these uncertainties on the slope
$\beta$ can then be explored using the Gaussian estimators
(\ref{eq:gauss}). These estimators have two advantages over the $\chi^2$ or AB
estimators for this purpose: (i) the slope estimator depends only on the
difference $\epsilon_y^2-\epsilon_x^2$ and hence is a function of only one
variable; (ii) the condition that the derived intrinsic dispersion
$\sigma_b^2$ be positive-definite provides an upper limit to the allowable
errors.

\begin{figure}
\plottwo{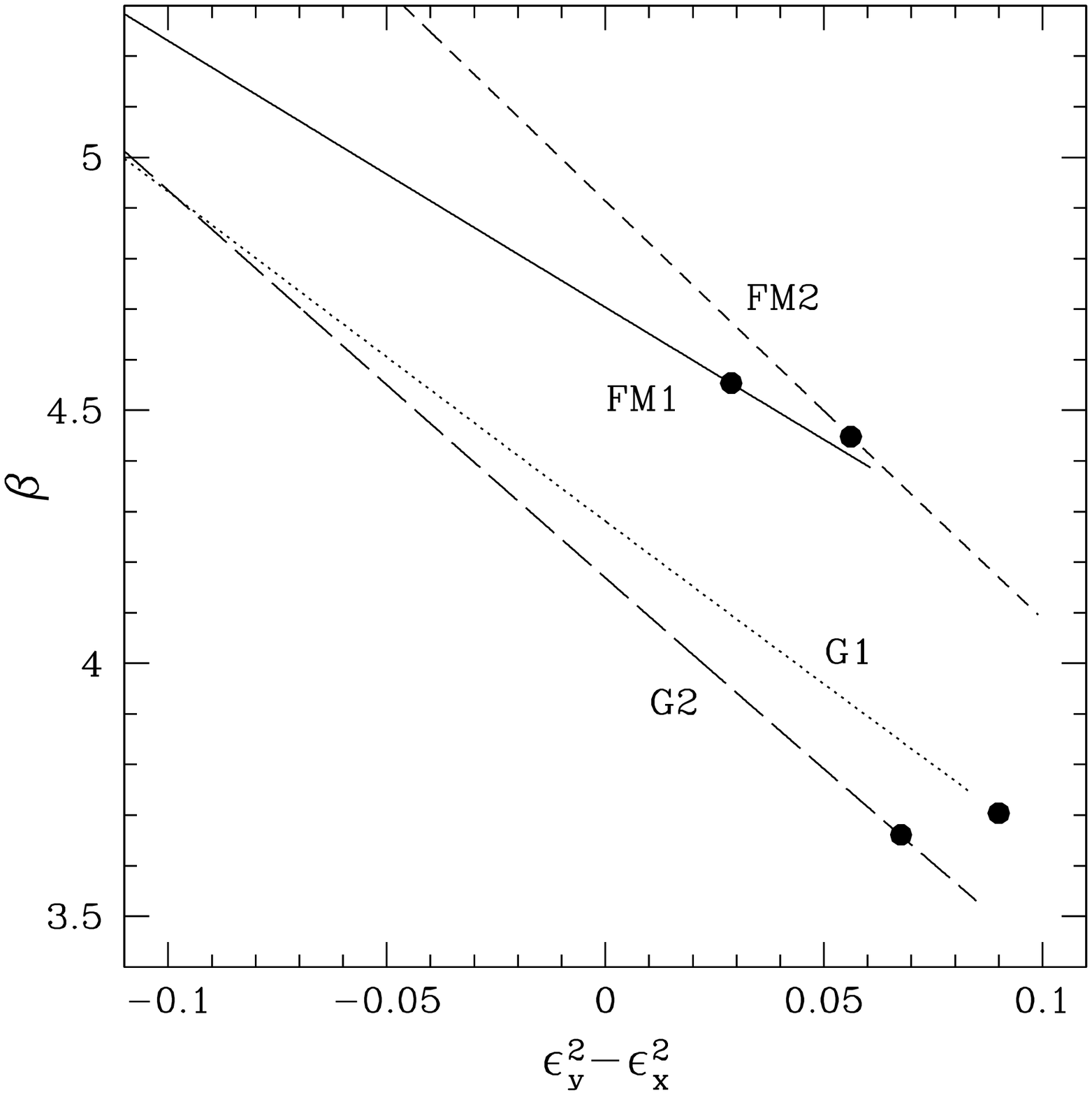}{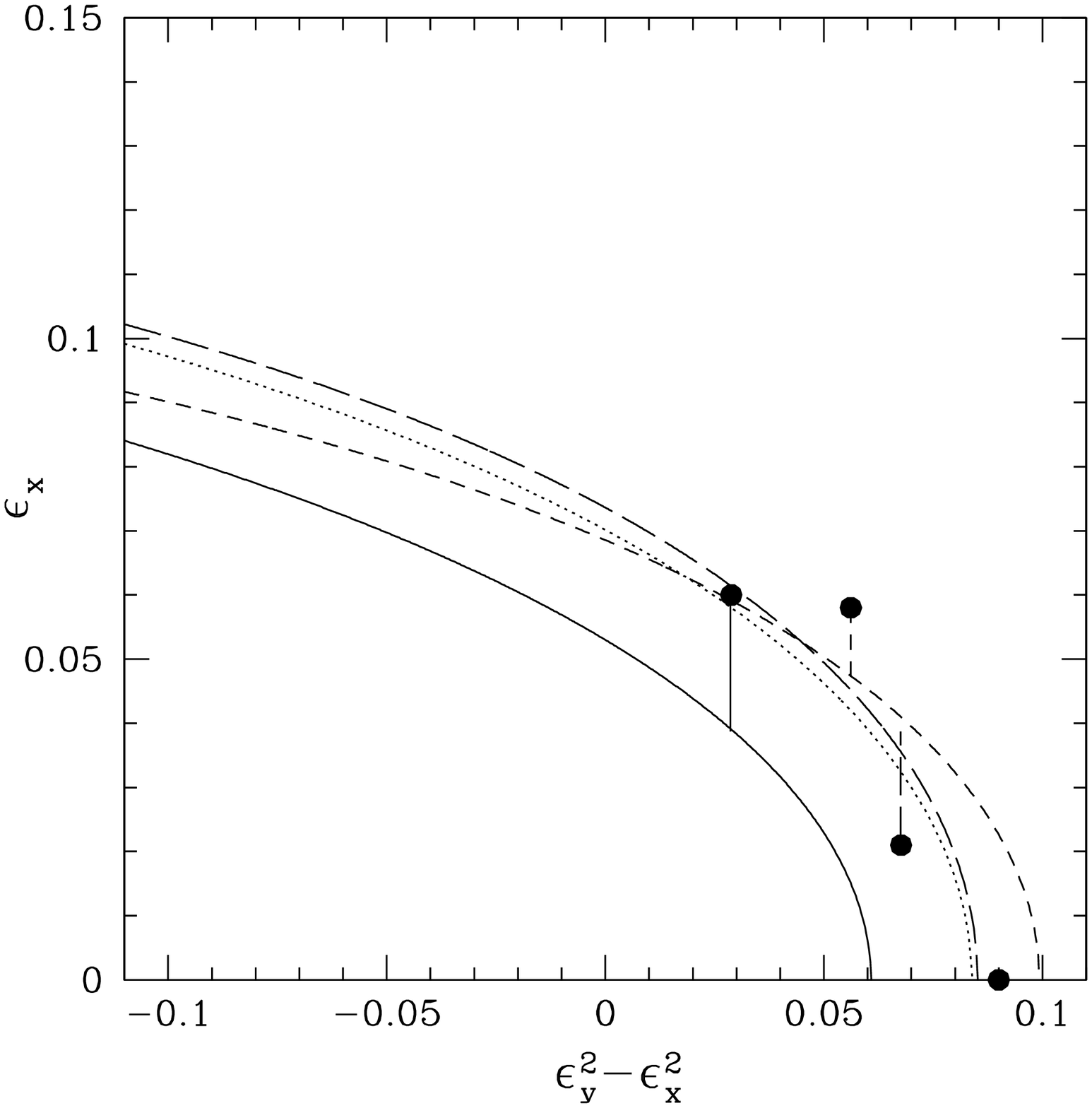}
\caption{The dependence of the slope $\beta$ on the assumed rms errors in
black-hole mass and velocity dispersion. The rms errors in
$\log M_\bullet$ and $\log\sigma$ are $\epsilon_y$ and $\epsilon_x$,
respectively (assumed the same for all galaxies). The left panel shows the
slope derived from the Gaussian estimator (\ref{eq:gauss}), for samples FM1
(solid line), FM2 (short-dashed line), G1 (dotted line), and G2 (long-dashed
line). The lines stop where the intrinsic dispersion $\sigma_b^2<0$. The right
panel shows the maximum allowed value of $\epsilon_x$; for larger values the
intrinsic dispersion is negative. The filled circles denote the locations
corresponding to the estimated values of $\epsilon_x$ and $\epsilon_y$ in each
survey; in the right panel these are connected by vertical lines to the curves
for the corresponding survey.}
\label{fig:be}
\end{figure}

The left and right panels of Figure \ref{fig:be} show the slope $\beta$ and
the maximum allowed value of $\epsilon_x$ for each of the galaxy samples in
\S\ref{sec:data}. For each sample there is a minimum slope $\beta$ and a
maximum value of $\epsilon_y^2-\epsilon_x^2$, beyond which the intrinsic
dispersion $\sigma_b^2$ is negative. In particular, for sample FM1 the minimum
allowable slope is $\beta=4.39$; thus there are {\em no} assumptions about the
measurement errors that can lead to a slope in the low range. The slope vs.\
error lines in the left panel of Figure \ref{fig:be} are approximately
parallel for all four samples; thus there is no single set of measurement
errors that could remove the discrepancy between the high slopes found by FM
and the low slopes found by the Nukers. Consistent slopes would require that
$(\epsilon_y^2-\epsilon_x^2)_{\rm Nuker}\simeq
(\epsilon_y^2-\epsilon_x^2)_{\rm FM}-0.07$. This relation, combined with the
constraint $\sigma_b^2>0$, cannot be satisfied with any plausible combination
of measurement errors---note in particular that $\epsilon_y$ should be similar
for the two groups since they rely on many of the same black-hole mass
determinations, and $\epsilon_x$ should be {\em smaller} for the Nuker samples
than the FM samples, since the Nukers employ high signal-to-noise ratio slit
spectra while FM rely on central velocity dispersions from the pre-1990
literature.  We conclude that random measurement errors cannot explain the
slope discrepancy.

\subsection{The dispersion of the Milky Way}

\label{sec:mw}

\begin{figure}
\plottwo{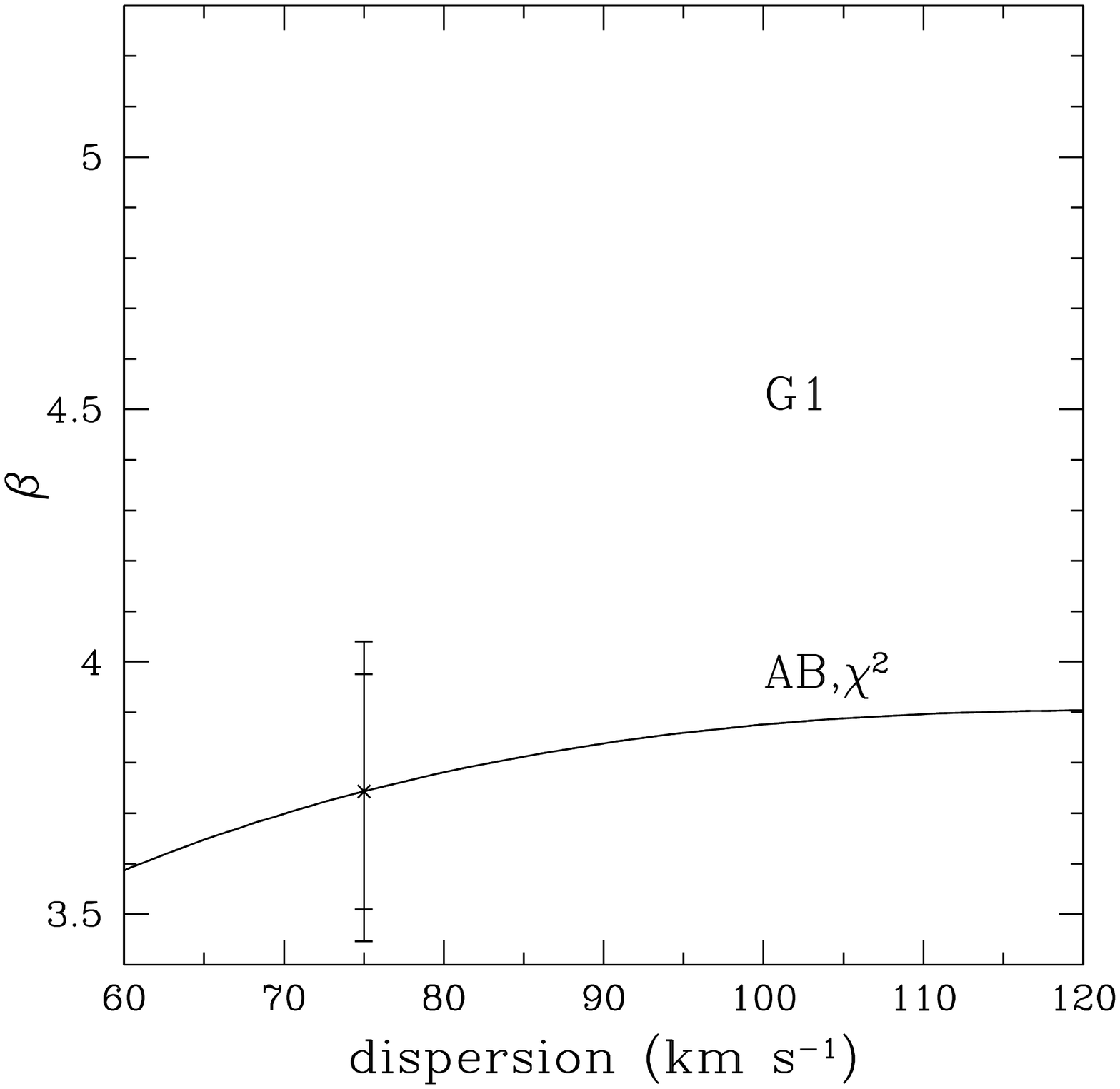}{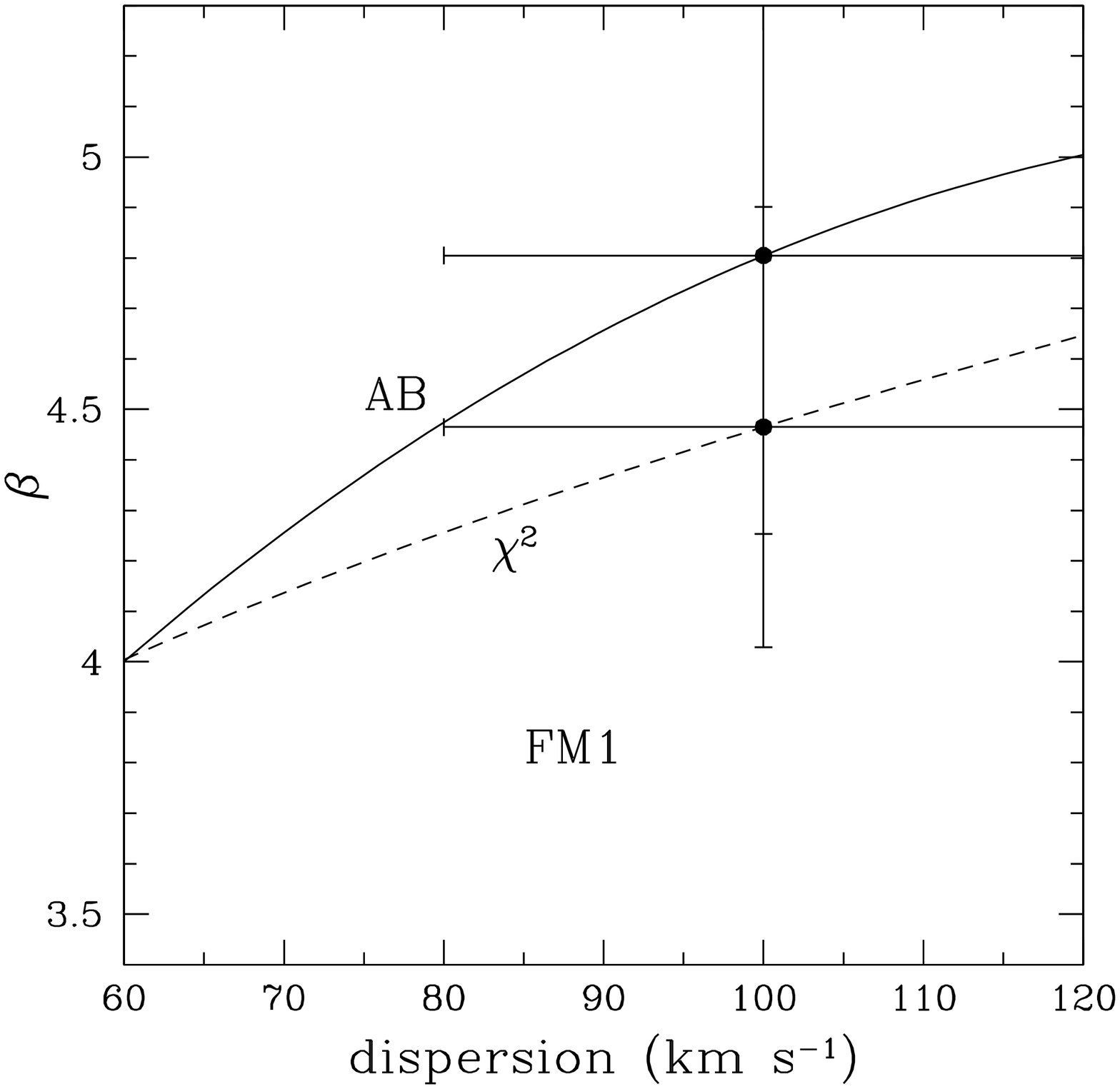}
\caption{The dependence of the slope $\beta$ on the assumed velocity
dispersion for the Milky Way, in samples G1 (left panel) and FM1
(right panel). The filled circles and error bars show the assumed dispersion
and the corresponding slope and error bars. Solid and dashed lines show the
slopes derived from the AB and $\chi^2$ estimators, respectively; these are the
same for the G1 sample because $\epsilon_x=0$ and $\epsilon_y$ is the same for
all galaxies.}
\label{fig:mw}
\end{figure}

\citet{MF01a} also argue that the slope is strongly affected by
the assumed dispersion for the Milky Way Galaxy, for which the Nukers'
estimated dispersion $\sigma=75\kms$ should be increased to
$\sigma=100\kms$. We show in Figure \ref{fig:mw} how the derived slope depends
on the Milky Way dispersion, for both samples G1 and FM1. We see that in fact
$\beta$ is quite insensitive to the Milky Way dispersion used in the G1
sample: increasing the dispersion from $75\kms$ to $100\kms$ as suggested by
\citet{MF01a} increases $\beta$ only by 0.13. The corresponding
slope change is substantially larger for sample FM1---$0.27$ for the $\chi^2$
estimator and $0.44$ for the AB estimator---but this strong sensitivity
reflects the small size of that sample and is not relevant to conclusions
drawn by \citet{G00} from sample G1.

Despite this conclusion, it is worthwhile to determine a more accurate value
for the Milky Way dispersion to use in the $M_\bullet$--$\sigma$ relation. We
review the data on the dispersion of the Galactic bulge in the Appendix, where
our results are summarized in the dispersion profile of Figure \ref{fig:disp}
and equation (\ref{eq:dispfit}). We stress that the dispersion profile of the
Milky Way is determined from a heterogeneous set of tracers with uneven
spatial coverage, and by very different methods than the dispersions of the
external galaxies discussed in this paper. We therefore assign our estimates
of the Milky Way dispersion an uncertainty of 20\%, much larger than the
formal uncertainty, and much larger than the 5\% uncertainty that we assume
for the dispersions of external galaxies.

The conversion of the dispersion profile in equation (\ref{eq:dispfit}) to a
characteristic dispersion is different for FM and the Nukers. FM define their
dispersion to be the luminosity-weighted rms line-of-sight dispersion within a
circular aperture of radius $r_e/8$, where $r_e$ is the effective radius. For
$r_e=0.7\kpc$ as derived in the Appendix, we find $\sigma=95\kms$. Because the
bulge is triaxial, we correct the dispersion that we measure from our
particular location to the average over all azimuths in the Galactic
plane. \citet{bin97} model the bulge as a triaxial system with axis ratios
1\,:\,0.6\,:\,0.4 and long axis at an angle $\phi_0=20^\circ$ from the
Sun-center line. If the density is stratified on similar ellipsoids, the ratio
$r^2\equiv \sigma^2(\phi_0=20^\circ)/\langle \sigma^2(\phi_0)\rangle$ depends
only on the axis ratios \citep{rob62}. For the axis ratios given by Binney et
al., $r=1.07$. Thus our best estimate for the dispersion within $r_e/8$ is
$\sigma_{\rm FM}=90\pm18\kms$; if we use this instead of FM's estimate of
$\sigma=100\pm20\kms$, the slope derived from sample FM1 is reduced from
$\beta_{\rm AB}=4.81\pm0.55$ to $\beta_{\rm AB}=4.66\pm0.42$, and for sample
FM2 from $\beta_{\rm AB}=4.65\pm 0.49$ to $4.54\pm 0.40$.

In contrast, the Nukers use the luminosity-weighted rms line-of-sight
dispersion within a slit aperture of half-length $r_e$. This dispersion
depends weakly on the slit width, which we take to be 70 pc (corresponding to
1 arcsec at Virgo). In this case we find $\sigma=110\kms$; reducing this by a
factor $r$ to account for triaxiality, we have $\sigma=103\pm20\kms$, close to
the value advocated by FM. This change increases the slope derived by
\citet{G00} from $\beta_\chi=3.74\pm0.15$ only to
$\beta_\chi=3.88\pm0.15$. Thus, improved estimates of the velocity dispersion
of the Milky Way bulge reduce the slope discrepancy only slightly.

\subsection{Different samples}

\label{sec:samp}

\citet{MF01b} argue that the shallower slope obtained by the Nukers
arises in part from the inclusion of galaxies in which the black-hole sphere of
influence is not well resolved. However, the samples FM2 and G2 contain exactly
the same 22 galaxies, all of which are claimed by \citet{MF01b} to have a
well-resolved sphere of influence, and the difference in slope $\beta_\chi$ 
(eqs.\ \ref{eq:mfb} and \ref{eq:kg}) is actually {\it larger} than between the
samples FM1 and G1.

\subsection{Aperture and effective dispersions}

\label{sec:apdisp}

\begin{figure}
\plotone{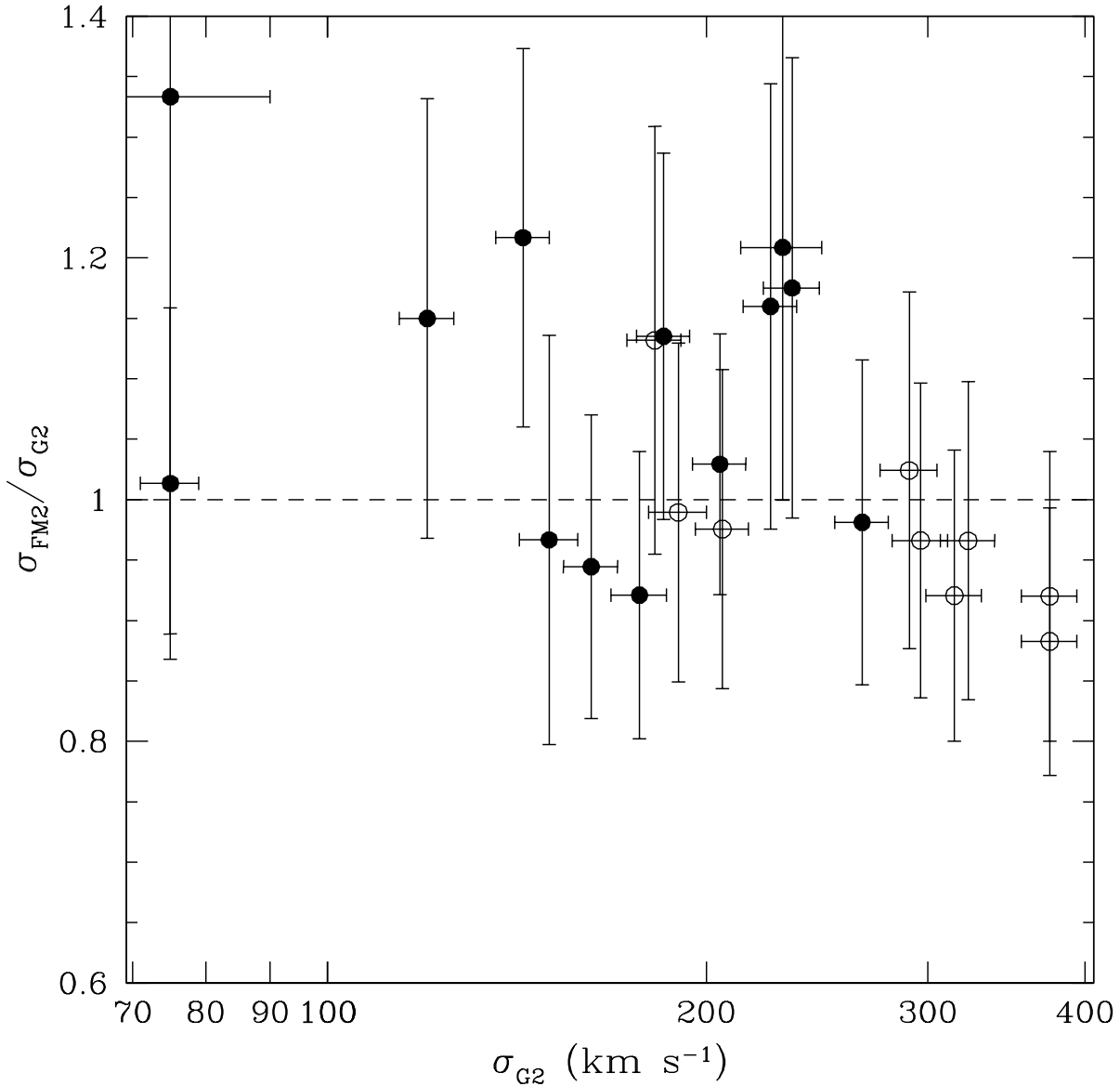}
\caption{Comparison of the velocity dispersions in the samples FM2 and G2. The
dispersion ratio $\sigma_{\rm FM2}/\sigma_{\rm G2}$ is plotted against
$\sigma_{\rm G2}$. Solid circles denote power-law galaxies and open circles
denote core galaxies. We take the uncertainties in the FM2 dispersions from
\citet{MF01b}, and assume that the uncertainties in the G2 dispersions are
20\% for the Milky Way and 5\% for external galaxies. The uncertainties in
dispersion ratio are computed by assuming that the errors in the FM2
dispersions and the G2 dispersions are independent. The plot shows that FM's
dispersions are higher than Nuker dispersions at low dispersion, and lower at
high dispersion.}
\label{fig:faber}
\end{figure}

Why, then, are the slopes different, particularly in the samples FM2 and G2,
which contain the same galaxies? If we fit the dispersions in these samples to
a relation of the form
\be
\log\sigma_{\rm G2}=\gamma+(1+\delta)\log\sigma_{\rm FM2},
\label{eq:ddd}
\ee
we find
\be
\delta_\chi=0.13\pm0.10,\qquad \delta_{\rm AB}=0.23\pm0.10,
\label{eq:deldef}
\ee
significantly different from the value $\delta=0$ that should obtain if there
were no systematic differences between the dispersions (see Figure
\ref{fig:faber}). A relation of this kind implies that the slopes $\beta$
determined from the FM2 and G2 samples will be related by $\beta_{\rm
FM2}/\beta_{\rm G2}=1+\delta$. Then if $\delta$ is in the range 0.15--0.20,
and the Nuker sample gives $\beta=4$, the sample from FM will give
$\beta=4.6$--4.8, well inside the high range. Thus it appears that the major
cause of the range of slopes is systematic differences in the dispersions:
FM's dispersions lead to high slopes, and the Nukers' dispersions lead to low
slopes.

This possibility was suggested by \citet{G00}, but was later rejected by
\citet{MF01a}, who argued that systematic differences between dispersions are
unimportant because there was ``remarkably little difference on average''
between the dispersions (they quote a mean ratio of 1.01 and a correlation
coefficient in the logs of 0.97). However, these statistics have no 
bearing on the slope $1+\delta$ in equation (\ref{eq:ddd}).

There are several possible explanations of the difference in dispersions:

\begin{enumerate}

\item The Nukers use the rms dispersion within a slit aperture of length
$2r_e$ (hereafter $\sigma_1$), while FM's results are based on the rms
dispersion within a circular aperture of radius $r_e/8$ (hereafter
$\sigma_8$). The ratio $\sigma_8/\sigma_1$ could depend systematically on the
velocity dispersion of the galaxy, for example,
\be
\sigma_1\propto \sigma_8^{1+\delta_1}.
\label{eq:delone}
\ee
In this case the difference in slope estimates would reflect the structural
properties of the galaxies. 

\item FM do not actually measure $\sigma_8$. Instead, they use the central
velocity dispersion (hereafter $\sigma_c$), typically measured in an aperture
of radius $r_{\rm ap}\simeq 2$ arcsec \citep{dav87}, and correct this to a
circular aperture of radius $r_e/8$ using the relation \citep{jor95}
$\sigma_8'=\sigma_c(8r_{\rm ap}/r_e)^{0.04}$ (the prime is used to distinguish
this approximation to $\sigma_8$ from the actual value of $\sigma_8$). The
ratio $\sigma_8'/\sigma_8$ could depend systematically on the velocity
dispersion of the galaxy, for example,
\be
\sigma_8\propto (\sigma_8')^{1+\delta_2}.
\label{eq:deltwo}
\ee
Such a trend could arise if the correction factor depends on galaxy luminosity
or type. In this case the difference in slope estimates would reflect
a shortcoming in FM's analysis rather than a real physical effect.

\item The dispersion measurements used by one or both of the two groups could
be subject to dispersion-dependent systematic errors (e.g., one set of
measurements is systematically low at high dispersions).  In this case the
difference in slope estimates would reflect problems with the data reduction. 

\end{enumerate}

To explore these possibilities, we have examined a sample of 40 early-type
galaxies for which \citet{fab97} have compiled effective radii and central
velocity dispersions, and have fitted {\it HST} photometry to a five-parameter
``Nuker law'' profile. Each galaxy is assumed to contain a central black hole,
with mass given by the $M_\bullet$--$\sigma$ relation in the form derived
below (eq.\ \ref{eq:chiff}). We use the Nuker law and the assumptions of
spherical symmetry, constant mass-to-light ratio, isotropic velocity
dispersion, and 1 arcsec slit width to compute the ratios $\sigma_8/\sigma_1$
and $\sigma_8'/\sigma_8$. This approach is model-dependent, but has the
advantages that (i) the discussion is independent of observational errors in
the dispersions, since the dispersion ratios are determined by a dynamical
model; (ii) the sample is larger, since more galaxies have {\it HST}
photometry than dispersion profiles. We find
\begin{eqnarray}
\log\left(\sigma_8\over\sigma_1\right)&=&(0.004\pm0.002)
+(0.021\pm0.010)\log\left(\sigma_c\over200\kms\right),\nonumber \\
\log\left(\sigma_8'\over\sigma_8\right)&=&-(0.012\pm0.003)
-(0.056\pm0.014)\log\left(\sigma_c\over200\kms\right).
\label{eq:drift}
\end{eqnarray}

The first of these equations suggests that there is a small but significant
systematic trend in the ratio $\sigma_8/\sigma_1$ with dispersion, of the form
(\ref{eq:delone}) with $\delta_1\simeq-0.02$. However, this trend has the
wrong sign and only a small fraction of the amplitude required to explain the
systematic differences in equations (\ref{eq:ddd}) and (\ref{eq:deldef}); thus
explanation 1 in the list above does not appear to be important.

The second of these equations suggests that there is a larger systematic trend
in the ratio $\sigma_8'/\sigma_8$ with dispersion, of the form
(\ref{eq:deltwo}) with $\delta_2\simeq 0.06$. This trend is sufficient to
explain about one-third of the systematic differences seen in equations
(\ref{eq:ddd}) and (\ref{eq:deldef}).  The origin of this trend, explanation 2
in the list above, is clear. \citet{geb96} and \citet{fab97} show that the
shape of the surface-brightness profile in the central parts of early-type
galaxies depends on the galaxy luminosity (and hence on its velocity
dispersion). Thus, the use of a single empirical formula to correct from
$\sigma_c$ to $\sigma_8$ will lead to systematic errors that are correlated
with velocity dispersion. It is always better to use the actual kinematic
observations, as was done in G1, than to apply empirical correction factors.

Explanation 3, dispersion-dependent systematic measurement errors, is more
difficult to assess. \citet{hud01} compare dispersion measurements from 27
sources, including the catalog used by FM \citep{dav87}, and in most cases
find no evidence for dispersion-dependent errors of the amplitude found in
equation (\ref{eq:deldef}) (see Hudson et al.'s Figure 3).  Nevertheless, it
is striking that the data points in Figure \ref{fig:faber} appear sharply
lower for dispersions $\gtrsim 300\kms$ than for smaller
dispersions. Measuring large dispersions is particularly difficult because the
spectral lines blend together. The principal conclusion is that we badly need a
systematic campaign of accurate {\it HST} and ground-based measurements of the
radial velocity-dispersion profiles of early-type galaxies with black-hole
candidates. A second conclusion is that the slope of the $M_\bullet$--$\sigma$
relation should be estimated only from dispersion measurements at or within
well-defined metric radii, rather than central velocity dispersions measured
within apertures of a given angular radius.

\begin{figure}
\epsscale{0.9}
\plotone{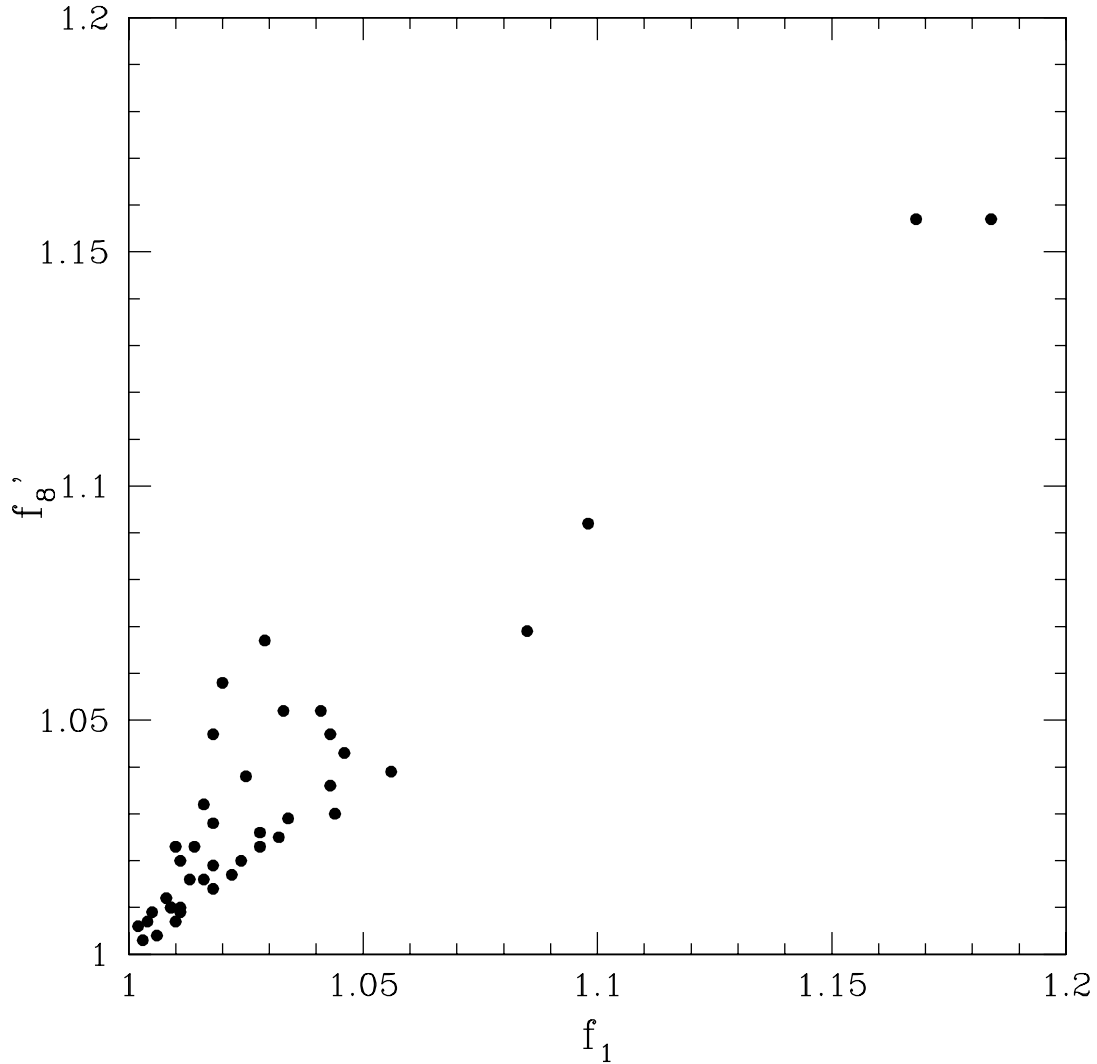}
\caption{Effect of central black holes on measured dispersions. For 40
early-type galaxies listed in \citet{fab97} we have computed isotropic,
spherical dynamical models that match the surface-brightness distributions and
mass-to-light ratios given in that paper. For each galaxy two models are
computed: one with no central black hole, and one with a black hole of mass
given by equation (\ref{eq:chiff}). The abscissa represents the ratio of the
dispersions $\sigma_1$ in these two models, where $\sigma_1$ is the
luminosity-weighted rms line-of-sight dispersion within a slit aperture of
half-length $r_e$ and width 1 arcsec (used by the Nukers). The ordinate
represents the ratio of the dispersions $\sigma_8'$; here $\sigma_8'$ is an
approximation to the dispersion within a circular aperture of radius $r_e/8$,
obtained from the central dispersion using an empirical correction formula
(used by FM). The galaxies in the upper right corner are M32 and NGC 3115.}
\label{fig:bhnobh}
\end{figure}

The same sample of 40 galaxies can also be used to explore the degree of
contamination of the dispersions by the dynamical influence of the central
black hole. We computed spherical, isotropic dynamical models with and without
a black hole of mass given by equation (\ref{eq:chiff}). We denote the ratio
of the dispersion $\sigma_1$ with and without the black hole by $f_1$, and the
analogous ratio for the dispersion $\sigma_8'$ by $f_8'$. The results are
shown in Figure \ref{fig:bhnobh}. In most galaxies the addition of the black
hole raises either dispersion no more than 3--4\%. However, in a few cases the
contamination is much larger, more than 15\%. In such galaxies the dispersion
measures $\sigma_c$, $\sigma_1$, $\sigma_8$, or $\sigma_8'$ are all
misleading. Future versions of the $M_\bullet$--$\sigma$ relation should be
based on dispersion measures that are less strongly weighted to the center.

\section{Black-hole mass vs. velocity dispersion: a new estimate}

\label{sec:new}

In this section we present a new analysis of the $M_\bullet$--$\sigma$
relation using the 31 galaxies in Table \ref{tab:mbh}; over half of these
have new or revised black-hole mass or dispersion determinations since the
analysis by \citet{G00}.

\begin{deluxetable}{lllllllll}
\tabletypesize{\footnotesize}
\tablecolumns{9}
\tablewidth{0pt}
\tablenum{1}
\tablecaption{Galaxy sample}
\tablehead{
\colhead{Galaxy} & \colhead{Type} & \colhead{$M_B$} & \colhead{$M_\bullet$ (low,high)} &
\colhead{Method} & \colhead{$\sigma_1$} & \colhead{Distance} & \colhead{$M/L$,} &
\colhead{Ref}\\
\colhead{} &\colhead{} & \colhead{} & \colhead{$M_\odot$} & \colhead{}
&\colhead{$\hbox{km s}^{-1}$} & \colhead{Mpc} & \colhead{band} &\colhead{}}
\startdata
Milky Way &SBbc & $-$17.65 & $1.8\times 10^6 (1.5,2.2)$ & s,p & 103 &   0.008& 1.0,K &  1 \\
N221=M32  & E2  & $-$15.83 & $2.5\times 10^6 (2.0,3.0)$ & s,3I&  75 &   0.81 &1.85,I &  5 \\
N224=M31  & Sb  & $-$19.00 & $4.5\times 10^7 (2.0,8.5)$ & s   & 160 &   0.76 & 5,V   &  2,3,4 \\
N821      & E4  & $-$20.41 & $3.7\times 10^7 (2.9,6.1)$ & s,3I& 209 &  24.1  & 5.8,V &  6,7  \\
N1023     & SB0 & $-$18.40 & $4.4\times 10^7 (3.9,4.9)$ & s,3I& 205 &  11.4  & 5.0,V &  8 \\
N1068     & Sb  & $-$18.82 & $1.5\times 10^7 (1.0,3.0)$ & m   & 151 &  15.0  &\nodata & 9 \\
N2778     & E2  & $-$18.59 & $1.4\times 10^7 (0.5,2.2)$ & s,3I& 175 &  22.9  &6.4,V & 6,7 \\ 
N2787     & SB0 & $-$17.28 & $4.1\times 10^7 (3.6,4.5)$ & g   & 140 &   7.5  &\nodata & 10 \\
N3115     & S0  & $-$20.21 & $1.0\times 10^9 (0.4,2.0)$ & s   & 230 &   9.7  &6.9,V    & 11 \\
N3245     & S0  & $-$19.65 & $2.1\times 10^8 (1.6,2.6)$ & g   & 205 &  20.9  & 3.7,R &  12 \\
N3377     & E5  & $-$19.05 & $1.0\times 10^8 (0.9,1.9)$ & s,3I& 145 &  11.2  & 2.7,V &  6,13 \\
N3379     & E1  & $-$19.94 & $1.0\times 10^8 (0.5,1.6)$ & s,3I& 206 &  10.6  & 4.6,V &  14 \\
N3384     & S0  & $-$18.99 & $1.6\times 10^7 (1.4,1.7)$ & s,3I& 143 &  11.6  & 2.8,V &  6,7 \\
N3608     & E2  & $-$19.86 & $1.9\times 10^8 (1.3,2.9)$ & s,3I& 182 &  22.9  & 3.7,V &  6,7 \\
N4258     & Sbc & $-$17.19 & $3.9\times 10^7 (3.8,4.0)$ & m,a & 130 &   7.2  &\nodata & 15 \\
N4261     & E2  & $-$21.09 & $5.2\times 10^8 (4.1,6.2)$ & g   & 315 &  31.6  & 5.0,V &  16 \\
N4291     & E2  & $-$19.63 & $3.1\times 10^8 (0.8,3.9)$ & s,3I& 242 &  26.2  & 4.4,V &  6,7 \\
N4342     & S0  & $-$17.04 & $3.0\times 10^8 (2.0,4.7)$ & s,3I& 225 &  15.3  & 6.3,I &  17 \\
N4459     & S0  & $-$19.15 & $7.0\times 10^7 (5.7,8.3)$ & g   & 186 &  16.1  &\nodata & 10 \\
N4473     & E5  & $-$19.89 & $1.1\times 10^8 (0.31,1.5)$& s,3I& 190 &  15.7  & 6.3,V & 6,7 \\
N4486=M87 & E0  & $-$21.53 & $3.0\times 10^9 (2.0,4.0)$ & g   & 375 &  16.1  & 4.0,V & 18,19 \\
N4564     & E3  & $-$18.92 & $5.6\times 10^7 (4.8,5.9)$ & s,3I& 162 &  15.0  & 1.9,I & 6,7 \\
N4596     & SB0 & $-$19.48 & $7.8\times 10^7 (4.5,12)$  & g   & 152 &  16.8  &\nodata & 10 \\
N4649     & E1  & $-$21.30 & $2.0\times 10^9 (1.4,2.4)$ & s,3I& 385 &  16.8  & 9.0,V & 6,7 \\
N4697     & E4  & $-$20.24 & $1.7\times 10^8 (1.6,1.9)$ & s,3I& 177 &  11.7  & 4.8,V & 6,7 \\
N4742     & E4  & $-$18.94 & $1.4\times 10^7 (0.9,1.8)$ & s,3I&  90 &  15.5  &\nodata & 20 \\
N5845     & E3  & $-$18.72 & $2.4\times 10^8 (1.0,2.8)$ & s,3I& 234 &  25.9  & 4.8,V & 6 \\
N6251     & E2  & $-$21.81 & $5.3\times 10^8 (3.5,7.0)$ & g   & 290 &  93.0  & 8.5,V &  21 \\
N7052     & E4  & $-$21.31 & $3.3\times 10^8 (2.0,5.6)$ & g   & 266 &  58.7  & 6.3,I &  22 \\
N7457     & S0  & $-$17.69 & $3.5\times 10^6 (2.1,4.6)$ & s,3I&  67 &  13.2  & 3.4,V & 6,7 \\
IC1459    & E3  & $-$21.39 & $2.5\times 10^9 (2.1,3.0)$ & s,3I& 340 &  29.2  &3.1,I &  23 \\
\enddata
\tablecomments{Distances are taken from \citet{ton01} for most of the
galaxies; where these are not available the distance is determined from the
recession velocity, assuming a Hubble constant of $80\kms\Mpc^{-1}$. Absolute
magnitudes are for the hot component of the galaxy only. The mass-to-light
ratios $M/L$ are usually determined from the same dynamical models that are
used to derive the black-hole masses; they are given here for reference but
play no role in our analysis. Methods: s=stellar radial velocities; p=stellar
proper motions; m=maser radial velocities; a=maser accelerations; g=rotating
gas disk from emission-line observations; 3I=axisymmetric dynamical models,
including three integrals of motion. References for the black-hole masses: 
(1) \citet{cs01}; (2)
\citet{tre95}; (3) \citet{kb99}; (4) \citet{bac01}; (5) \citet{ver02}; (6)
\citet{geb02}; (7) \citet{pin02}; (8) \citet{bow01}; (9) 
\citet{gre97}; (10) \citet{sar00}; (11) \cite{kor96a}; 
(12) \citet{bar01}; (13) \citet{kor98}; (14)
\citet{G00b}; (15) \citet{her99}; (16) \citet{fer96}; (17) \citet{cre99}; (18)
\citet{har94}; (19) \citet{mac97}; (20) \citet{kai01};
(21) \citet{fer99}; (22) \citet{vdm98b}; (23) \cite{cap02}}
\label{tab:mbh}
\end{deluxetable}

\subsection{Comments on individual galaxies}

{\em Milky Way:} We use the black-hole mass estimate by \citet{cs01},
$(1.8{+0.4\atop-0.3})\times10^6M_\odot$. For comparison, \citet{ghez98} find
$(2.4\pm0.2)\times10^6M_\odot$, and \citet{gen00} find
(2.6--3.3)$\times10^6M_\odot$. The dispersion and its uncertainty are
discussed in \S \ref{sec:mw}. The bulge mass-to-light ratio is taken from
\citet{ken92}. 

{\em M32:} The velocity dispersion is obtained from \citet{vdm94} and the
black-hole mass, $M_\bullet=(2.5\pm0.5)\times10^6M_\odot$, is from
\citet{ver02}. In estimating the dispersion, we have excluded the region near
the center that is strongly perturbed by the black hole (cf. Figure
\ref{fig:bhnobh}). Other recent mass estimates, by
\citet{vdm98a} and \citet{jos01}, give similar results: $(3.9\pm0.8)\times
10^6M_\odot$ and $(3\pm 1)\times10^6M_\odot$, respectively.

{\em M31:} The modeling is complicated by the double nucleus. \citet{kb99}
find $M_\bullet=(3.0\pm1.5)\times10^7M_\odot$, although this result relies
heavily on the small displacement between the center of light of the nucleus
and bulge. \citet{tre95} and \citet{bac01} find $M\simeq 7\times10^7M_\odot$,
but without detailed model fitting. We adopt the range
(2.0--8.5)$\times10^7M_\odot$.

{\em NGC 1068:} The black-hole mass is taken from \citet{gre97}; the error
estimates are our own and are very approximate. The dispersion \citep{kob00} is
somewhat uncertain because of contamination from the bright nucleus.

{\em NGC 3115:} The black-hole mass is based on stellar kinematics
\citep{kor96a}. Although NGC 3115 does not have three-integral axisymmetric
dynamical models, it does have a compact, high-contrast stellar nucleus, and
the mass of the nucleus plus black hole can be estimated from the virial
theorem. In estimating the dispersion, we have excluded the region near
the center that is strongly perturbed by the black hole (cf. Figure
\ref{fig:bhnobh}).

{\em NGC 3245:} The velocity dispersion is obtained from \citet{sim98}.

{\em NGC 4258:} The velocity dispersion is obtained from \citet{her98}. 

{\em NGC 4486:} The mass is the average of the values given by \citet{har94}
and \citet{mac97}, corrected to a distance of 16.1 Mpc. 

{\em NGC 2787, NGC 4459, NGC 4596:} The black-hole masses are based on Space
Telescope Imaging Spectrograph (STIS) measurements of ionized-gas disks by
\citet{sar00}. The disk inclinations are determined from dust-lane
morphology. Note that the distance and dispersion for NGC 2787, 7.5 Mpc and
$140\kms$, are much smaller than the values assumed by \citet{sar00}. Our
distance is from \citet{ton01} and the dispersion was measured by one of us
(Gebhardt). For NGC 4459 and NGC 4596 we have used the dispersions $\sigma_8'$
from \cite{sar00}, since on average these are close to $\sigma_1$ (eq.\
\ref{eq:drift}). 

{\em IC 1459:} The mass estimate that we use \citep{cap02}, based on stellar
kinematics, is much larger than an earlier estimate by the same group from gas
kinematics, (2--6)$\times10^8M_\odot$ \citep{vk00}. The mass estimate from
stellar kinematics is much more reliable, since the gas rotation curve is
asymmetric and non-Keplerian. 

We do not include the following galaxies in our sample:

NGC 4594 \citep{kor96b}, NGC 4486B \citep{kor97}, NGC 4350 \citep{pig01}, NGC
3031=M81, and NGC 3998 \citep{bow00} exhibit strong evidence from 
stellar dynamics for a black hole, but do not yet have three-integral
dynamical models.

NGC 4374 (M84) has strong evidence for a black hole from gas dynamics, but the
published estimates of the black-hole mass differ by far more than the stated
errors: \citet{bow98} find (0.9--2.6)$\times10^9M_\odot$; \citet{mb01} find
$4\times10^8M_\odot$, and \citet{bar01a} find $10^9M_\odot$. The mass assigned
to this galaxy is a factor of four larger in sample FM2 than in sample G2,
which is by far the largest discrepancy between the two samples.

NGC 4945 has a mass estimate from maser emission \citep{gre97b} but no reliable
dispersion.

NGC 5128 has a mass estimate from ground-based observations of a rotating gas
disk \citep{mar01} but no {\it HST} spectroscopy; moreover, the galaxy has
peculiar morphology, presumably because of a recent merger, and thus may not
follow the same $M_\bullet$--$\sigma$ relation as more normal galaxies. 

Our sample contains eight galaxies with black-hole mass estimates based on gas
kinematics. We have some concern that these results may have large systematic
errors, due in part to uncertainties in the the spatial distribution of the
gas (e.g., filled disk or torus configuration; uncertain inclination and
thickness) and the large but uncertain correction for pressure support. We
therefore urge caution when interpreting results from samples in which a large
fraction of the black-hole mass estimates are based on gas
kinematics. Eventually, galaxies with black-hole mass determinations from more
than one technique will be invaluable for disentangling the systematic errors
in different methods.

\subsection{Slope estimation}

\label{sec:slope}

We use the sample of galaxies and black-hole masses in Table \ref{tab:mbh} to
estimate the logarithmic slope $\beta$ in the $M_\bullet$--$\sigma$
relation. We assume 20\% uncertainty in the dispersion of the Milky Way
(cf. \S\ref{sec:mw}) and 5\% uncertainties in the dispersions of external
galaxies (cf. \S\ref{sec:disperr}), although the uncertainties in the
dispersions of a few galaxies that we have not observed ourselves may be
larger. Initially we assume 0.33 dex rms uncertainties in the black-hole
masses, which yields $\chi^2$ per degree of freedom of unity. Using the
$\chi^2$ and AB estimators defined in \S\ref{sec:fit}, we find
\be
\beta_\chi=4.03\pm0.33, \qquad \beta_{\rm AB}=4.12\pm0.34.
\label{eq:chifa}
\ee

\begin{figure}
\plotone{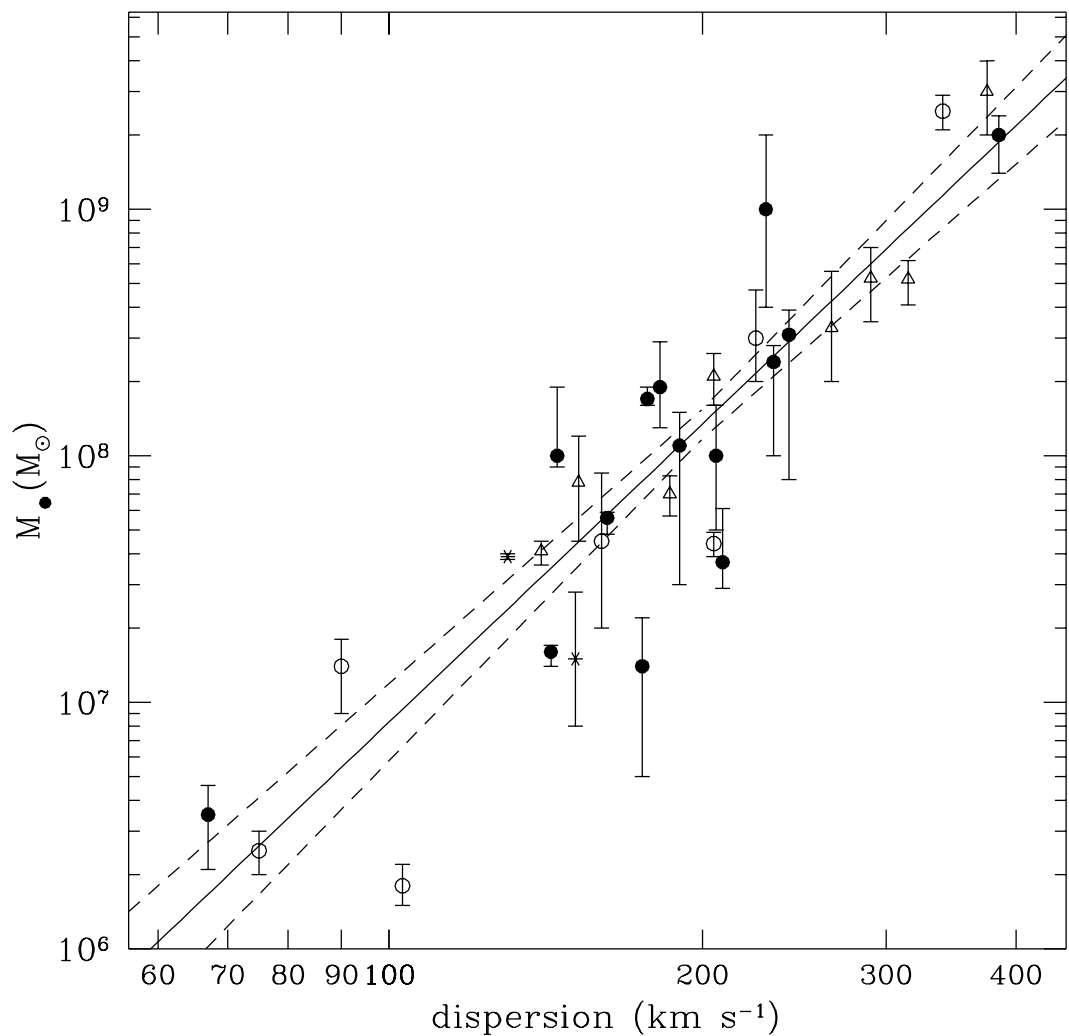}
\caption{The data on black-hole masses and dispersions for the galaxies in
Table \ref{tab:mbh}, along with the best-fit correlation described by
equations (\ref{eq:gebrel}) and (\ref{eq:chiff}). Mass
measurements based on stellar kinematics are denoted by circles, on gas
kinematics by triangles, and on maser kinematics by asterisks; Nuker
measurements are denoted by filled circles. The dashed lines show the
1$\sigma$ limits on the best-fit correlation.}
\label{fig:msig}
\end{figure}

This approach does not account for the varying precision of the mass estimates
for different galaxies. Therefore we have also computed the slope using the
estimated errors in the black-hole masses in Table \ref{tab:mbh}, adding to
these in quadrature a common intrinsic dispersion with rms value $\epsilon_0$
(i.e., $\epsilon_{yi}\to(\epsilon_{yi}^2+\epsilon_0^2)^{1/2}$). We find that
$\epsilon_0=0.27$ gives
\be
\beta_\chi=4.00\pm0.31, \qquad \beta_{\rm AB}=4.12\pm0.34,
\label{eq:chifb}
\ee
with minimum $\chi^2$ per degree of freedom of 1.00. A maximum-likelihood
estimate of the intrinsic dispersion in black-hole mass at constant velocity
dispersion for this sample is $\epsilon_0=0.23\pm0.05$ dex.

In both equations (\ref{eq:chifa}) and (\ref{eq:chifb}) the AB estimator for
the slope is larger than the $\chi^2$ estimator by about 0.1; since we have
shown in \S\ref{sec:fit} that the AB estimator may be biased, we prefer to
rely on the $\chi^2$ estimator. For our final answer we simply average
$\beta_\chi$ from equations (\ref{eq:chifa}) and (\ref{eq:chifb}). Including
results for the parameter $\alpha$ obtained in the same way, we have
\be
\alpha=8.13\pm0.06, \qquad \beta=4.02\pm0.32; 
\label{eq:chiff}
\ee
the parameter $\alpha$ is evaluated for $\sigma_0=200\kms$, for which the
correlation coefficient between $\alpha$ and $\beta$ is only $-0.09$. 

Thus our best estimate (\ref{eq:chiff}) is just at the edge of the low range,
$\beta\lesssim 4.0$. To test the robustness of this result, we have tried
culling the sample in several ways:

\begin{itemize}

\item If we consider only the 21 galaxies from Table \ref{tab:mbh} with masses
determined from stellar kinematics, we find
\be
\alpha=8.13\pm 0.09, \qquad \beta=4.02\pm 0.44;
\label{eq:chiffs}
\ee
the close agreement in the parameters in equations (\ref{eq:chiff}) and
(\ref{eq:chiffs}) implies that there is no significant systematic bias between
masses determined by stellar kinematics and other methods.

\item The dispersions for the Milky Way and for external galaxies are
determined by quite different methods. The Milky Way also has one
of the smallest and most accurate black-hole masses in our sample, and
therefore has an unusually strong influence on the slope of the
$M_\bullet$--$\sigma$ relation. If we remove the Milky Way from our sample,
the slope is reduced to $\beta=3.88\pm0.32$, a change of 0.14 (0.4 standard
deviations) towards even lower slopes.  

\item We have argued in \S\ref{sec:apdisp} that high velocity-dispersion
measurements may be subject to systematic errors. Thus we also estimate the
slope using only the 25 galaxies in the sample with dispersion less than
$250\kms$. We find $\beta=3.77\pm0.49$; once again the slope is even lower
than our best estimate (\ref{eq:chiff}).

\item The galaxy sample with the most homogeneous observations and analysis
consists of the 10 galaxies analyzed by \citet{pin02} and \citet{geb02}. These
all have {\it HST} spectra acquired with STIS as well as ground-based spectra,
{\it HST} photometry, and axisymmetric orbit-based dynamical models, and were
all reduced and analyzed in the same way. For this sample we find
$\beta=3.67\pm0.70$; once again the slope is consistent with and even lower
than our best estimate.

\item We have removed 9 galaxies from the sample which were subject to
criticism: the Milky Way (uncertain dispersion), M31 (no accurate models of
the double nucleus), NGC 1068 (both the dispersion and the interpretation of
the maser kinematics are uncertain), NGC 2778 (the lowest signal-to-noise
ratio in the Gebhardt et al. 2002 sample and a correspondingly large
uncertainty in the black-hole mass), NGC 3115 (no three-integral dynamical
models), NGC 3379 and NGC 5845 (these have only a single FOS pointing rather
than STIS slit spectra at {\it HST} resolution; while there is no obvious
problem with either measurement, other galaxies in the Gebhardt et al. 2002
sample have superior spatial coverage of the kinematics), NGC 4459 (the
inclination of the gas disk is uncertain because the kinematic data comes from
a single long-slit spectrum; also, the dispersion is uncertain because it is
obtained from low-resolution data), and NGC 6251 (the most distant galaxy with
a black-hole mass measurement; the sphere of influence of the black hole is
poorly resolved and in addition there are the usual uncertainties---uncertain
disk orientation, influence of random motions in the gas---associated with
mass measurements from gas kinematics). The reduced sample of 22 galaxies has
a slope $\beta=3.79\pm0.32$, once again lower than our best estimate.

\end{itemize}

Since most of these culled samples have slopes that are smaller than our best
fit (\ref{eq:chiff}), we suspect that our best fit may slightly
overestimate the true slope by 0.1--0.3.

\begin{figure}
\plotone{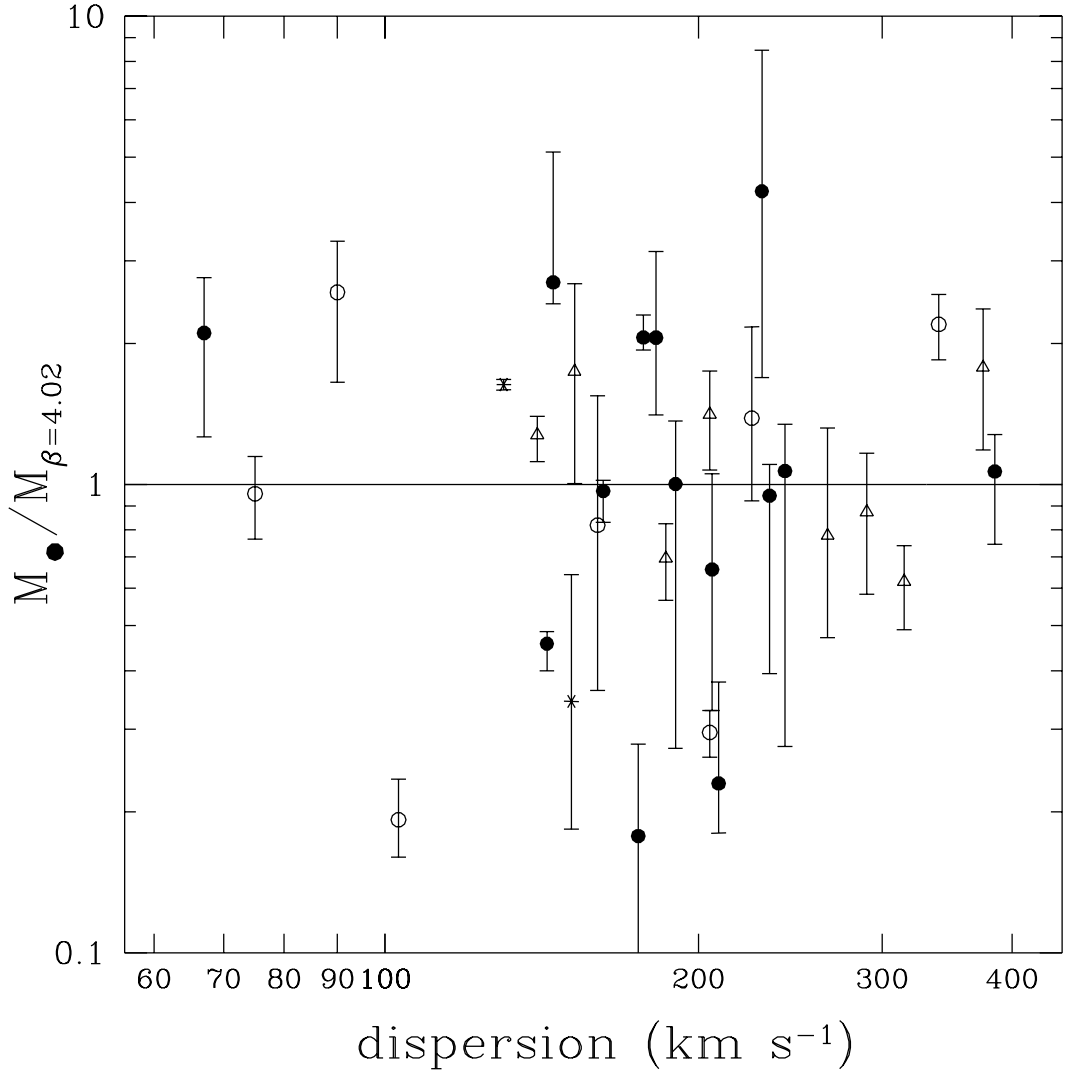}
\caption{The residuals between the black-hole masses and dispersions for the
galaxies in Table \ref{tab:mbh} and the best-fit correlation described by
equation (\ref{eq:gebrel}) with $\beta=4.02$ (eq.\ \ref{eq:chiff}).  Mass
measurements based on stellar kinematics are denoted by circles, on gas
kinematics by triangles, and on maser kinematics by asterisks; Nuker
measurements are denoted by filled circles.}
\label{fig:resid}
\end{figure}

The data from Table \ref{tab:mbh} and the fit (\ref{eq:chiff}) are shown in
Figure \ref{fig:msig}. In Figure \ref{fig:resid} we show the residuals to the
best-fit correlation.

The two largest residuals in Figure \ref{fig:msig} belong to NGC 2778 ($-0.75$
dex) and the Milky Way ($-0.72$ dex); the largest positive residual belongs to
NGC 3115 ($+0.63$ dex). The poor fit of the Milky Way probably arises because
its dispersion profile has been determined by different methods than the other
galaxies, using heterogeneous tracers and a variety of surveys; we have
allowed for this by assigning the Milky Way dispersion an uncertainty of 20\%,
compared to 5\% for external galaxies. The large residual in NGC 3115 may
arise because its mass has been estimated by simply applying the virial
theorem to its nucleus, rather than by dynamical modeling. The large residual
in NGC 2778 may reflect the low signal-to-noise ratio of its kinematic data
\citep{pin02}. 

\section{Conclusions}

The masses $M_\bullet$ of dark objects (``black holes'') in the centers of
nearby early-type galaxies are related to the velocity dispersion $\sigma$ by
the log-linear relation (\ref{eq:gebrel}). We have used the sample of 31
galaxies in Table \ref{tab:mbh} to determine the parameters in this relation,
where $\sigma$ is defined to be the luminosity-weighted rms velocity
dispersion in a slit extending to the effective radius. Our best estimate for
the slope of this relationship is $4.0\pm0.3$ (eq.\ \ref{eq:chiff}), although
several culled, and perhaps higher quality, samples give slopes that are lower
by 0.1--0.3. There is no evidence for systematic differences in either slope
or normalization between black-hole mass measurements based on stellar
kinematics and gas kinematics.  If the stated measurement errors in the
black-hole masses are correct, or if they are underestimated because of
systematic errors, the intrinsic dispersion in the $M_\bullet$--$\sigma$
relation is no larger than about 0.3 dex in black-hole mass (i.e., less than a
factor of two).

The range of slopes for the $M_\bullet$--$\sigma$ relation found in the
literature appears to arise mostly from systematic differences in the velocity
dispersions used by different groups. We do not believe that these differences
reflect the different definitions of dispersion used by the groups (FM use the
dispersion within a circular aperture of radius $r_e/8$, and the Nukers use
the dispersion within a slit aperture of half-length $r_e$). It appears that
part of the difference results from Ferrarese \& Merritt's analysis, in which
central velocity dispersions are extrapolated to $r_e/8$ using an empirical
formula. However, another---and possibly larger---component appears to arise
from poorly understood systematic errors in the dispersion measurements.

In a few galaxies, the influence of the central black hole may significantly
affect the velocity dispersions---both the central dispersions used by FM and
the slit dispersions used by the Nukers. Future analyses of the
$M_\bullet$--$\sigma$ relation should be based on velocity-dispersion measures
that are less strongly weighted to the center. Other improvements in the
analysis would include the use of statistical estimators that are more robust
and that explicitly include an intrinsic dispersion in the black-hole mass,
accounting properly for the asymmetric error bars in black-hole mass
determinations, and estimating more accurately the uncertainties in individual
dispersion measurements.

The investment of the astronomy community in the difficult task of measuring
black-hole masses has not yet been matched by a commensurate investment in the
much easier task of obtaining high-quality kinematic maps of galaxies
containing black holes. A complete set of high-quality dispersion and rotation
profiles for the galaxies in Table \ref{tab:mbh} would allow us to explore
more deeply how the black-hole mass is related to the kinematic structure
of its host galaxy.

\acknowledgments

We thank Michael Hudson and Tim de Zeeuw for discussions, and Tim de Zeeuw for
communicating results in advance of publication. Support for proposals 7388,
8591, 9106, and 9107 was provided by NASA through a grant from the Space
Telescope Science Institute, which is operated by the Association of
Universities for Research in Astronomy, Inc., under NASA contract NAS
5-26555. This research was also supported by NSF grant AST-9900316.

\appendix

\section{The effective dispersion for the Milky Way}

\label{app:mw}

The Milky Way has one of the most accurate black-hole masses and anchors the
low-mass end of the $M_\bullet$--$\sigma$ relation. Therefore it is important
to have an accurate value for the dispersion of the Milky Way bulge.

The first task is to estimate the effective or half-light radius $r_e$ of the
bulge. \citet{ken92} models Spacelab $K$-band observations of the bulge with a
major-axis emissivity profile of the form
\be
j(a)=\cases{j_ia^{-1.85}  & for $a<0.94\kpc$\cr
            j_oK_0(a/a_0) & for $a>0.94\kpc$,} 
\label{eq:kent}
\ee
where $K_0$ is a modified Bessel function, $a_0=0.67\kpc$, and the constants
$j_i$ and $j_o$ are chosen so that the emissivity is continuous. In a
spherical galaxy described by equation (\ref{eq:kent}), the effective radius
is $1.50a_0$ or 1.0 kpc; Kent's model is oblate and axisymmetric, with axis
ratio 0.6, so the geometric mean of the three effective semi-axes is smaller
by $(0.6)^{1/3}$, yielding $r_e=0.84$ kpc.

\citet{dwe95} fit COBE measurements in several bands to a wide variety
of triaxial models for the emissivity. Their best-fit model at $K$-band has a
Gaussian emissivity profile with an effective semi-major axis of 1.86 kpc; the
axis ratios are 1\,:\,0.4\,:\,0.3, so our best estimate for the effective
radius is $r_e=1.86\kpc\,(0.4\times 0.3)^{1/3}=0.92\kpc$. Their second-best
model (E3) has $j(a)\propto K_0(a/a_0)$ and an effective radius
$r_e=0.56\kpc$.

\citet{bin97} use COBE $L$-band photometry to perform a
disk/\-bulge decomposition. Their equation (1b) describes an analytic
model for the bulge emissivity that fits the data ``very well'':
\be 
j(a)=j_0{e^{-a^2/a_m^2}\over(1+a/a_0)^{1.8}}, 
\label{eq:binn} 
\ee
where $a$ is the semi-major axis and $a_m=1.9$ kpc. They quote $a_0=100$ pc,
but this value reflects the fact that the data have been smoothed to an angular
resolution of 1.5$^\circ$ or 200 pc, and photometry at higher
resolution suggests that $a_0$ is less than 1 pc (e.g., Genzel et
al.\ 1996). The effective semi-major axis for equation (\ref{eq:binn}) is
$0.48a_m$ or $0.91\kpc$; the corresponding geometric mean of the effective
semi-axes  (1\,:\,0.6\,:\,0.4) is $r_e=0.57\kpc$.

Based on these estimates, we shall adopt $r_e=0.7\pm0.2\kpc$. The much larger
estimate $r_e=2.7\kpc$ given by \citet{MF01a} is based on a table in
\citet{gil90}, which in turn appears to be based on
the galaxy model of \citet{bs80}, which in turn is based on comments by G.\ de
Vaucouleurs in the 1970s that $r_e$ is about one-third of the distance
of the Sun from the Galactic Center.

The next task is to estimate the velocity dispersion as a function of radius.
We are interested in the rms line-of-sight velocity $\langle v_{\rm
los}^2\rangle^{1/2}$ measured relative to the Local Standard of Rest, since
this is the closest analog to the dispersions used in the
$M_\bullet$--$\sigma$ relation for external galaxies. This quantity differs
from the usual dispersion quoted in bulge studies, which is relative to the
local mean velocity, $\sigma=\langle(v_{\rm los}-\overline v)^2\rangle^{1/2}$,
where $\overline v=\langle v_{\rm los}\rangle$. When papers quote values for
$\sigma$ and $\overline v$ we set $\langle v_{\rm
los}^2\rangle=\overline v^2+\sigma^2$. We use the following sources:

\begin{deluxetable}{lll}
\tablecolumns{3}
\tablewidth{0pt}
\tablenum{2}
\tablecaption{Velocity dispersion measurements in the inner bulge
($r< 1\kpc$)}
\tablehead{
\colhead{radius (pc)} & \colhead{$\langle v_{\rm los}^2\rangle^{1/2}$} & \colhead{Source}} 
\startdata
0.085 & $195\pm 34 $ &\citet{gen00} \\ 
0.33  & $164\pm 74 $ &\quad''\quad'' \\
0.34  & $102\pm 8  $ &\quad''\quad'' \\
0.39  & $ 99\pm 10 $ &\quad''\quad'' \\
0.67  & $ 72\pm 5 $ &\quad''\quad'' \\
0.78  & $ 85\pm 15 $ &\quad''\quad'' \\
1.2   & $ 68\pm 13 $ &\quad''\quad'' \\
3.9   & $ 54\pm  6 $ &\quad''\quad'' \\
15.3  & $ 70\pm  7 $ &\citet{lind92a} \\
38.5  & $101\pm 11 $ &\quad''\quad'' \\
117   & $126\pm 14 $ &\citet{seven97} \\
160   & $156\pm 18 $ &\citet{blum95} \\
171   & $128\pm 14 $ &\quad''\quad'' \\
288   & $129\pm 14 $ &\quad''\quad'' \\
299   & $148\pm 19 $ &\quad''\quad'' \\
314   & $130\pm 14 $ &\citet{seven97} \\ 
527   & $101\pm 11 $ &\quad''\quad'' \\
562   & $110\pm 10 $ &\citet{tern95} \\ 
612   & $117\pm 12 $ &\citet{beau99} \\
789   & $ 88\pm  9 $ &\quad''\quad'' \\
851   & $102\pm 12 $ &\citet{seven97} \\
989   & $100\pm 10 $ &\citet{beau99} \\
1220  & $ 89\pm  9 $ &\quad''\quad'' \\
1284  & $ 79\pm  8 $ &\citet{seven97} \\
\enddata
\label{tab:mw}
\end{deluxetable}

\begin{enumerate}

\item Due to the interest in the black hole in our Galaxy, the kinematics in
the central few parsecs have been investigated much more thoroughly than the
kinematics at larger radii (Genzel et al.\ 2000, especially their Figure 16).
The entries at radii $<5\pc$ in Table \ref{tab:mw} are taken from Genzel et
al.'s Table 4; at these radii corrections for rotation are negligible.

\item OH/IR stars are mass-losing asymptotic giant branch stars, which are
detected by hydroxyl maser emission from their circumstellar envelopes. They
are old enough to represent a phase-mixed population and are unaffected by
obscuration, and hence should be good tracers of the kinematics of the
bulge. The survey by Lindqvist et al.\ (1992a,b) lists 133 OH/IR stars within
$1^\circ$ or 140 pc of the Galactic Center. We have divided these into three
equal groups by projected distance from the center and computed the dispersion
for each group. One limitation of this survey is that its radial-velocity
coverage was relatively small, $|v_{\rm los}|\le 217\kms$, so that
high-velocity OH/IR stars might have been missed. We have corrected for this
cutoff, assuming that the distribution of line-of-sight velocities is
Gaussian, in the two bins where the correction is less than 10\%, and have
discarded the third bin. At larger distances, \citet{seven97} have
located 307 OH/IR stars in the region $|\ell|<10^\circ$, $|b|<3^\circ$. The
minimum velocity range in this survey was $-330\kms<v_{\rm los}<402\kms$ so
velocity selection effects are negligible. We have discarded all sources not
having a standard double-peaked profile and all sources with expansion
velocity $>17\kms$, which appear to represent a younger, more rapidly rotating
population \citep{wi98}. The remaining 208 stars were divided into five equal
groups by projected distance, and the mean projected distance and dispersion
were computed for each group.
  
\item \citet{beau99} have conducted an H$\alpha$ survey for new planetary 
nebulae and remeasured the velocities of many known planetary nebulae. Their
databases contain 183 planetary nebulae within $10^\circ$ of the Galactic
Center. We have divided these into four equal groups by projected distance,
and the mean projected distance and dispersion were computed for each
group. Beaulieu et al. estimate that their velocity errors are $\pm11\kms$,
which is negligible.

\item Blum et al.\ (1994, 1995) have measured the dispersion of samples of M
giants in four fields between 160 and 300 pc from the Galactic
Center. \citet{tern95} have measured the dispersion of K giants in Baade's
window (0.56 kpc from the Galactic Center). We include only stars with
$V>16.0$, which they believe restricts the sample to bulge stars and
eliminates the foreground disk.

\end{enumerate} 

\begin{figure}
\plotone{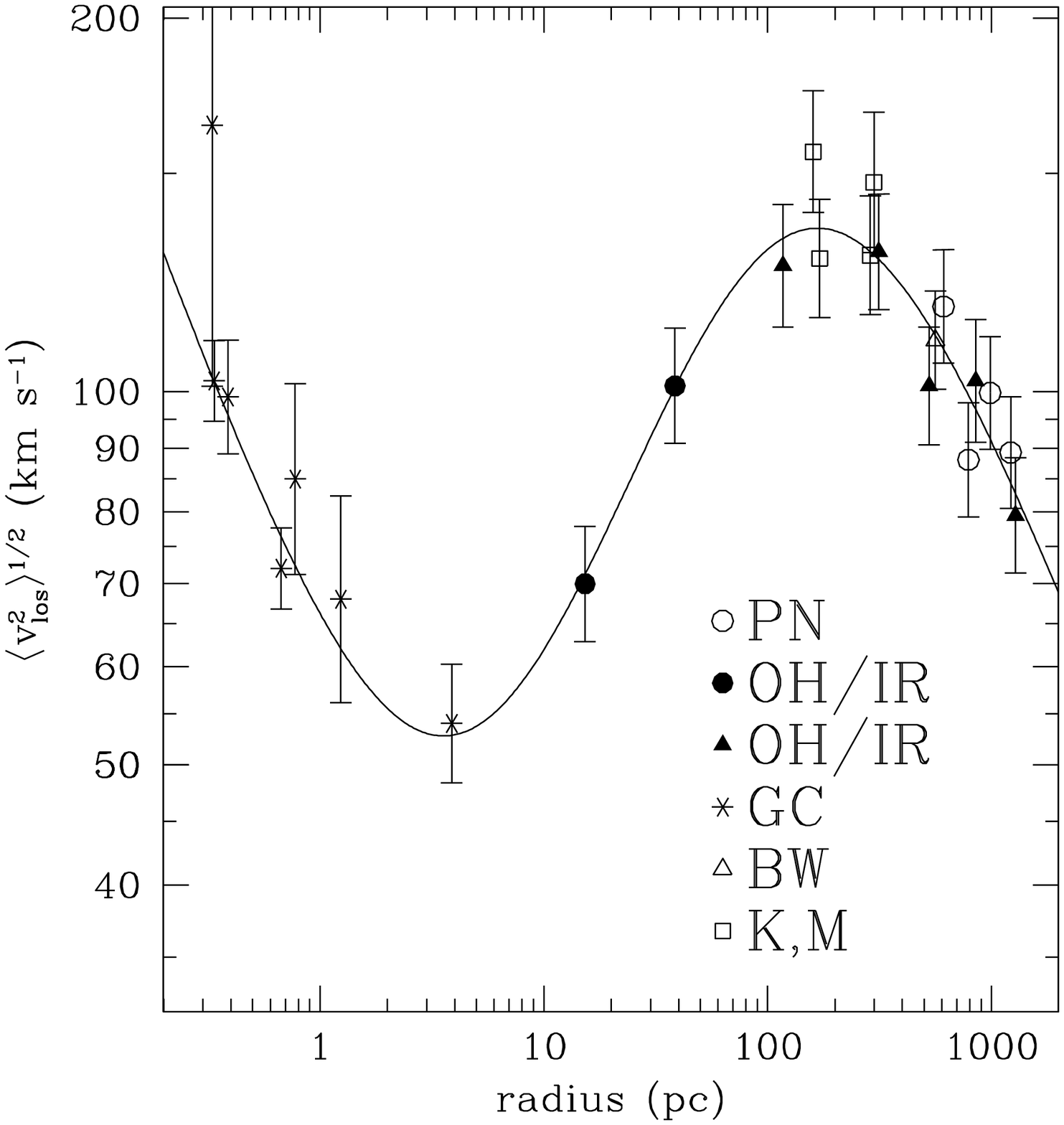}
\caption{The rms line-of-sight velocity in the bulge of the Milky Way, as a
function of radius. PN = planetary nebulae \citep{beau99}; OH/IR = OH/IR stars
\citep{lind92a,lind92b,seven97}; BW = giant stars in Baade's window
\citep{tern95}; K,M = giant stars \citep{blum94,blum95}; GC = stars near the
Galactic Center \citep{gen00}. Filled symbols denote observations biased
toward the Galactic plane, and open symbols denote observations biased away
from the plane. The curve is the fitting function (\ref{eq:dispfit}). }
\label{fig:disp}
\end{figure}

The data at $r>0.1\pc$ shown in Figure \ref{fig:disp} and Table \ref{tab:mw}
have been fit to the functional form
\be
\langle v_{\rm los}^2\rangle^{1/2}=c{(r/r_0)^\alpha\over 1+b(r/r_0)^\beta}+
d(r/r_0)^{-1/2}.
\label{eq:dispfit}
\ee
For $r_0=500\pc$ the best-fit values are $c=633\kms$, $\alpha=0.67$,
$\beta=1.14$, $b=4.64$, $d=2.52$. The general features of this curve---a
minimum in the dispersion near 5 pc and a maximum of $\sim 130\kms$ at
a few hundred pc---are not new \citep{ken92}.

A possible concern is that the bulge is flattened, with an axis ratio of about
0.5, so the dispersion at a given radius may depend on the angle between the
radius vector and the Galactic plane. To address this concern, we have divided
the data points from outside 4 pc from the Galactic Center into those biased
towards the minor axis, plotted with filled symbols (the criterion is
$\langle|\ell|\rangle > \langle|b|\rangle$, where $\ell$ and $b$ are the
Galactic longitude and latitude; these are objects such as planetary nebulae
and late-type giants that are found optically), and those biased towards the
major axis (the OH/IR stars, found in surveys along the Galactic plane), which
are plotted with open symbols. There is no obvious systematic difference
between the dispersion curves defined by the filled and open symbols.

We employ the fit (\ref{eq:dispfit}) at the end of \S\ref{sec:mw} to estimate
the appropriate Milky Way dispersion to use in the $M_\bullet$--$\sigma$
relation.


\begin{thebibliography}{99}

\bibitem[Adams, Graff, \& Richstone(2001)]{ada01} Adams, F.\ C., Graff, D.\
S., \& Richstone, D.\ O.\ 2001, ApJ, 551, L31

\bibitem[Akritas \& Bershady(1996)]{ab96} Akritas, M.\ G., \& Bershady,
M.\ A.\ 1996, ApJ, 470, 706

\bibitem[Bacon et al.(2001)]{bac01} Bacon, R., Emsellem, E., Combes, F.,
Copin, Y., Monnet, G., \& Martin, P.\ 2001, A\&A, 371, 409

\bibitem[Bahcall \& Soneira(1980)]{bs80} Bahcall, J.\ N., \& Soneira, R.\ M.\
1980, ApJS, 44, 73

\bibitem[Barth et al.(2001a)]{bar01} Barth, A.\ J., Sarzi, M., Rix, H.-W., Ho,
L.\ C., Filippenko, A.\ V., \& Sargent, W.L.W.\ 2001a, ApJ, 555, 685

\bibitem[Barth et al.(2001b)]{bar01a} Barth, A.\ J., Sarzi, M., Ho, L.\ C., 
Rix, H.-W., Shields, J.\ C., Filippenko, A.\ V., Rudnick, L., \& Sargent,
W.L.W.\ 2001b, in The Central Kiloparsec of Starbursts and AGNs, eds.\ J.\ H.\
Knapen, J.\ K.\ Beckman, I.\ Shlosman, \& T.\ J.\ Mahoney (San
Francisco: Astronomical Society of the Pacific), in press (astro-ph/0110672) 

\bibitem[Beaulieu, Dopita, \& Freeman(1999)]{beau99} Beaulieu, S., Dopita,
M.\ A., \& Freeman, K.\ C.\ 1999, ApJ, 515, 610

\bibitem[Binney, Gerhard, \& Spergel(1997)]{bin97} Binney, J., Gerhard, O., \&
Spergel, D.\ 1997, MNRAS, 288, 365 

\bibitem[Blum et al.(1994)]{blum94} Blum, R.\ D., Carr, J.\ S., Depoy, D.\ L.,
Sellgren, K., \& Terndrup, D.\ M.\ 1994, ApJ, 422, 111

\bibitem[Blum et al.(1995)]{blum95} Blum, R.\ D., Carr, J.\ S., Sellgren,
K., \& Terndrup, D.\ M.\ 1995, ApJ, 449, 623

\bibitem[Bower et al.(1998)]{bow98} Bower, G.\ A., et al.\ 1998, ApJ, 492, L111

\bibitem[Bower et al.(2000)]{bow00} Bower, G.\ A., Wilson, A.\ S., Heckman,
T.\ M., Magorrian, J., Gebhardt, K., Richstone, D.\ O., Peterson, B.\ M., \&
Green, R.\ F.\ 2000, AAS meeting 197, 92.03

\bibitem[Bower et al.(2001)]{bow01} Bower, G.\ A., et al.\ 2001, ApJ, 550, 75

\bibitem[Burkert \& Silk(2001)]{bur01} Burkert, A., \& Silk, J.\ 2001, ApJ,
554, L151

\bibitem[Cappellari et al.(2002)]{cap02} Cappellari, M., et al. 2002,
astro-ph/0202155 

\bibitem[Chakrabarty \& Saha(2001)]{cs01} Chakrabarty, D., \& Saha, P.\ 2001,
AJ, 122, 232

\bibitem[Cretton \& van den Bosch(1999)]{cre99} Cretton, N., \& van den Bosch,
F.\ 1999, ApJ, 514, 704

\bibitem[Davies et al.(1987)]{dav87} Davies, R.\ L., Burstein, D., Dressler,
A., Faber, S.\ M., Lynden-Bell, D., Terlevich, R.\ J., \& Wegner, G.\ 1987,
ApJS, 64, 581

\bibitem[Dwek et al.(1995)]{dwe95} Dwek, E., et al.\ 1995, ApJ, 445, 716

\bibitem[Faber et al.(1997)]{fab97} Faber, S.\ M., et al.\ 1997, AJ, 114, 1771

\bibitem[Fabian \& Iwasawa(1999)]{fab99} Fabian, A., \& Iwasawa, K.\ 1999,
MNRAS, 303, L34

\bibitem[Ferrarese(2002)]{fer02} Ferrarese, L.\ 2002, in ``Current High-Energy
Emission around Black Holes'', ed.\ C.-H. Lee (Singapore: World Scientific),
in press 

\bibitem[Ferrarese \& Ford(1999)]{fer99} Ferrarese, L., \& Ford, H.\ C.\ 1999,
ApJ, 515, 583

\bibitem[Ferrarese \& Merritt(2000a)]{FM00a} Ferrarese, L., \& Merritt,
D.\ 2000a, astro-ph/0006053 v1

\bibitem[Ferrarese \& Merritt(2000b)]{FM00b} Ferrarese, L., \& Merritt,
D.\ 2000b, ApJ, 539, L9 (sample FM1)

\bibitem[Ferrarese, Ford, \& Jaffe(1996)]{fer96} Ferrarese, L., Ford, H.\ C.,
\& Jaffe, W.\ 1996, ApJ, 470, 444

\bibitem[Gebhardt et al.(1996)]{geb96} Gebhardt, K., et al.\ 1996, AJ, 112, 105

\bibitem[Gebhardt et al.(2000a)]{G00} Gebhardt, K., et al.\ 2000a, ApJ, 539,
L13 (sample G1)

\bibitem[Gebhardt et al.(2000b)]{G00b} Gebhardt, K., et al.\ 2000b, AJ, 119,
1157 

\bibitem[Gebhardt et al.(2002)]{geb02} Gebhardt, K., et al.\ 2002, in
preparation 

\bibitem[Genzel et al.(1996)]{gen96} Genzel, R., Thatte, N., Krabbe, A.,
Kroker, H., \& Tacconi-Garman, L.\ E.\ 1996, ApJ, 472, 153

\bibitem[Genzel et al.(2000)]{gen00} Genzel, R., Pichon, C., Eckart, A.,
Gerhard, O.\ E., \& Ott, T.\ 2000, MNRAS, 317, 348

\bibitem[Ghez et al.(1998)]{ghez98} Ghez, A.\ M., Klein, B.\ L., Morris, M., \&
Becklin, E.\ E.\ 1998, ApJ, 509, 678

\bibitem[Gilmore, King, \& van der Kruit(1990)]{gil90} Gilmore, G., King, I.\
R., \& van der Kruit, P.\ C.\ 1990, The Milky Way as a Galaxy (Mill Valley:
University Science Books) 

\bibitem[Greenhill \& Gwinn(1997)]{gre97} Greenhill, L.\ J., \& Gwinn, C.\ R.\
1997, Ap\&SS, 248, 261

\bibitem[Greenhill, Moran, \& Herrnstein(1997)]{gre97b} Greenhill, L.\ J.,
Moran, J.\ M., \& Herrnstein, J.\ R.\ 1997, ApJ, 481, L23

\bibitem[Gull(1989)]{gu89} Gull, S.\ F.\ 1989, in Maximum Entropy and Bayesian
Methods, ed. J.\ Skilling (Dordrecht: Kluwer), 511

\bibitem[Haehnelt \& Kauffmann(2000)]{hae00} Haehnelt, M.\ G., \& Kauffmann,
G.\ 2000, MNRAS, 318, L35

\bibitem[Harms et al.(1994)]{har94} Harms, R.\ J., et al.\ 1994, ApJ, 435, L35

\bibitem[H\'eraudeau \& Simien(1998)]{her98} H\'eraudeau, P., \& Simien, F.\
1998, A\&AS, 133, 317

\bibitem[Herrnstein et al.(1999)]{her99} Herrnstein, J.\ R., et al.\ 1999,
Nature, 400, 539 

\bibitem[Hudson et al.(2001)]{hud01} Hudson, M.\ J., Lucey, J.\ R., Smith, R.\
J., Schlegel, D.\ J., \& Davies, R.\ L.\ 2001, MNRAS, 327, 265

\bibitem[J\o rgensen, Franx, \& Kjaergaard(1995)]{jor95} J\o rgensen, I.,
Franx, M., \& Kjaergaard, P.\ 1995, MNRAS, 276, 1341

\bibitem[Joseph et al.(2001)]{jos01} Joseph, C.\ L., et al.\ 2001, ApJ, 550,
668 

\bibitem[Kaiser et al.(2002)]{kai01} Kaiser, M.\ E., et al.\ 2002, in
preparation 

\bibitem[Kendall, Stuart, \& Ord(1983)]{ken83} Kendall, M., Stuart, A., \& Ord,
J.\ K.\ 1983, The Advanced Theory of Statistics, 3, 4th ed. (London: Charles
Griffin) 

\bibitem[Kent(1992)]{ken92} Kent, S.\ M.\ 1992, ApJ, 387, 181

\bibitem[Kobulnicky \& Gebhardt(2000)]{kob00} Kobulnicky, H.\ A., \& 
Gebhardt, K.\ 2000, AJ, 119, 1608

\bibitem[Kormendy \& Bender(1999)]{kb99} Kormendy, J., \& Bender, R.\ 1999,
ApJ, 522, 772

\bibitem[Kormendy \& Gebhardt(2001)]{kg01} Kormendy, J., \& Gebhardt, K.\ 2001,
in The 20th Texas Symposium on Relativistic Astrophysics, eds.\ H.\ Martel \&
J.\ C.\ Wheeler (New York: AIP), in press (astro-ph/0105230) (sample G2)

\bibitem[Kormendy et al.(1996a)]{kor96a} Kormendy, J., et al.\ 1996a, ApJ,
459, L57 

\bibitem[Kormendy et al.(1996b)]{kor96b} Kormendy, J., et al.\ 1996b, ApJ,
473, L91 

\bibitem[Kormendy et al.(1997)]{kor97} Kormendy, J., et al.\ 1997, ApJ, 482,
L139 

\bibitem[Kormendy et al.(1998)]{kor98} Kormendy, J., Bender, R., Evans, A.\
S., \& Richstone, D.\ 1998, AJ, 115, 1823

\bibitem[Lindqvist et al.(1992a)]{lind92a} Lindqvist, M., Winnberg, A.,
Habing,  H.\ J., \& Matthews, H.\ E.\ 1992a, A\&AS, 92, 43

\bibitem[Lindqvist, Habing, \& Winnberg(1992b)]{lind92b} Lindqvist, M.,
Habing, H.\ J., \& Winnberg, A.\ 1992b, A\&A, 259, 118

\bibitem[Macchetto et al.(1997)]{mac97} Macchetto, F., Marconi, A., Axon, D.\
J., Capetti, A., Sparks, W., \& Crane, P. 1997, ApJ, 489, 579

\bibitem[Maciejewski \& Binney(2001)]{mb01} Maciejewski, W., \& Binney,
J.\ 2001, MNRAS, 323, 831

\bibitem[Marconi et al.(2001)]{mar01} Marconi, A., Capetti, A., Axon, D.\ J.,
Koekemoer, A., Macchetto, D., \& Schreier, E.\ J.\ 2001, ApJ, 549, 915

\bibitem[Merritt \& Ferrarese(2001a)]{MF01a} Merritt, D., \& Ferrarese,
L.\ 2001a, ApJ, 547, 140

\bibitem[Merritt \& Ferrarese(2001b)]{MF01b} Merritt, D., \& Ferrarese,
L.\ 2001b, in The Central kpc of Starbursts and AGN, eds. J.\ H.\ Knapen et
al. (San Francisco: Astronomical Society of the Pacific), in press
(astro-ph/0107134) (sample FM2)

\bibitem[Ostriker(2000)]{ost00} Ostriker, J.\ P.\ 2000, Phys. Rev. Lett. 84,
5258 

\bibitem[Pignatelli, Salucci, \& Danese(2001)]{pig01} Pignatelli, E., 
Salucci, P., \& Danese, L.\ 2001, MNRAS, 320, 124

\bibitem[Pinkney et al.(2002)]{pin02} Pinkney, J., et al.\ 2002, in
preparation 

\bibitem[Press et al.(1992)]{press92} Press, W.\ H., Teukolsky, S.\ A.,
Vetterling, W.\ T., \& Flannery, B.\ P.\ 1992, Numerical Recipes, 2nd
ed.\ (Cambridge: Cambridge University Press)

\bibitem[Roberts(1962)]{rob62} Roberts, P.\ H.\ 1962, ApJ, 136, 1108

\bibitem[Sarzi et al.(2001)]{sar00} Sarzi, M., Rix, H.-W., Shields, J.\ C.,
Rudnick, G., Ho, L.\ C., McIntosh, D.\ H., Filippenko, A.\ V., \& Sargent,
W.L.W.\ 2001, ApJ, 550, 65

\bibitem[Sevenster et al.(1997)]{seven97} Sevenster, M.\ N., Chapman, J.\ M.,
Habing, H.\ J., Killeen, N.E.B., \& Lindqvist, M.\ 1997, A\&AS, 122, 79

\bibitem[Simien \& Prugniel(1998)]{sim98} Simien, F., \& Prugniel, P.\ 1998,
A\&AS, 131, 287

\bibitem[So\l tan(1982)]{sol82} So\l tan, A.\ 1982, MNRAS, 200, 115

\bibitem[Terndrup, Sadler, \& Rich(1995)]{tern95} Terndrup, D.\ M., Sadler,
E.\ M., \& Rich, R.\ M.\ 1995, AJ, 110, 1774

\bibitem[Tonry et al.(2001)]{ton01} Tonry, J.\ L., Dressler, A., Blakeslee,
J.\ P., Ajhar, E.\ A., Fletcher, A.\ B., Luppino, G.\ A., Metzger, M.\ R., \&
Moore, C.\ B.\ 2001, ApJ, 546, 681

\bibitem[Tremaine(1995)]{tre95} Tremaine, S.\ 1995, AJ, 110, 628

\bibitem[van der Marel et al.(1994)]{vdm94} van der Marel, R.\ P., Rix, H.-W.,
Carter, D., Franx, M., White, S.D.M., \& de Zeeuw, T.\ 1994, MNRAS, 268, 521

\bibitem[van der Marel \& van den Bosch(1998)]{vdm98b} van der Marel, R.\ P.,
\& van den Bosch, F.\ C.\ 1998, AJ, 116, 2220

\bibitem[van der Marel et al.(1998)]{vdm98a} van der Marel, R.\ P., Cretton,
N., de Zeeuw, P.\ T., \& Rix, H.-W.\ 1998, ApJ, 493, 613

\bibitem[Verdoes Kleijn et al.(2000)]{vk00} Verdoes Kleijn, G.\ A., van der
Marel, R.\ P., Carollo, M., \& de Zeeuw, P.\ T.\ 2000, AJ, 120, 1221

\bibitem[Verolme et al.(2002)]{ver02} Verolme, E.\ K., et al.\ 2002, submitted
to MNRAS (astro-ph/0201086) 

\bibitem[Winnberg et al.(1998)]{wi98} Winnberg, A., Lindqvist, M., \& Habing,
H.\ J.\ 1998, in The Central Parsecs of the Galaxy, eds.\ H.\ Falcke et
al. (San Francisco: ASP), 389

\bibitem[Yu \& Tremaine(2002)]{yu02} Yu, Q., \& Tremaine, S.\ 2002,
astro-ph/0203082 


\end{thebibliography}
\end{document}